\documentclass[journal]{IEEEtran}
\usepackage{cite}
\usepackage{amsmath}
\usepackage{algorithm}
\usepackage{algpseudocode}
\usepackage{float}
\usepackage{graphicx,subfigure}
\usepackage{epstopdf}
\usepackage{amsmath,bm}
\usepackage{booktabs}
\usepackage{mathtools}
\usepackage{array}
\usepackage{tabularx}
\usepackage{filecontents,lipsum}
\usepackage{subfigure}
\usepackage{xcolor}
\DeclarePairedDelimiter\ceil{\lceil}{\rceil}

\begin{document}

	\title{ Effect of Transmission Impairments in    CO-OFDM Based Elastic Optical Network Design }
	\author{ Sadananda Behera, Jithin George, Goutam Das\\  
		
		\thanks{Sadananda Behera and Goutam Das are with the G. S. Sanyal School of Telecommunications, Indian Institute of Technology Kharagpur, India (email: sadanandabehera07@gmail.com, gdas@gssst.iitkgp.ernet.in). \par
			%		Arnab Mitra is with Microsoft India (R\&d) Private Limited, Hyderabad, India (email: arnabmitra08@gmail.com).
			
			Jithin George is with the Department of Electrical and Electronics Engineering, University of Melbourne,  Australia  (email: jithing@student.unimelb.edu.au).
		}}% <-this % stops a space
		
\maketitle

\IEEEpeerreviewmaketitle

\begin{abstract}
Coherent Optical Orthogonal Frequency Division Multiplexing (CO-OFDM) based Elastic Optical Network (EON) is one of the emerging technologies being considered for next generation high data rate optical network systems. Routing and Spectrum Allocation (RSA) is an important aspect of EON. Apart from spectral fragmentation created due to spectrum continuity and  contiguity constraints of RSA, transmission impairments  such as shot noise, amplified spontaneous emission  (ASE) beat noise  due to coherent detection, crosstalk in cross-connect (XC), nonlinear interference, and filter narrowing,  limit the transmission reach of optical signals in EON. This paper focuses on the cross-layer joint optimization of delay-bandwidth product, fragmentation and link congestion for RSA in CO-OFDM EON while considering the effect of physical layer impairments. First, we formulate an optimal Integer Linear Programming (ILP) that achieves load-balancing  in presence of transmission impairments and minimizes delay-bandwidth product along with fragmentation.  We next propose a heuristic algorithm for large networks with two different demand ordering techniques. We show the benefits of our algorithm compared to the existing load-balancing algorithm. 
\end{abstract}


\section{Introduction}
Driven by the increase in  demand  for high speed broadband, high definition video,  multimedia services, on-line gaming, etc., today's optical networks  are expected to  deploy data rates of the order of 100 Gbps  and beyond. 
The current wavelength-routed  optical  networks (WRONs) require full allocation of the wavelength capacity (fixed-grid allocation) to an end-to-end optical path even when the requested demand is not sufficient to use the whole wavelength capacity.
Due to this  rigid nature of fixed-grid allocation, the network-utilization efficiency in the current WRONs is limited. To combat this problem, recently, Elastic Optical Network (EON) is gaining  attention with its advanced spectrum-efficient, flexible, scalable, and adaptive features \cite{jinno2009spectrum}.  The main characteristics of EON include segmentation of larger demands into sub-wavelengths, aggregation of smaller demands to create super-wavelengths, efficient accommodation of multiple data rates, adaptive variation of allocated resources, etc. \cite{jinno2010distance}.  Coherent Optical Orthogonal Frequency Division Multiplexing (CO-OFDM) \cite{kobayashi2009over,sanjoh2002optical,sano2009no}  based EON, enabled by adaptive-allocation of network resources for varying traffic patterns, is the new trend in the field of core optical networking. 
Similar to Routing and Wavelength Assignment (RWA) in Wavelength Division Multiplexing (WDM), Routing and Spectrum Assignment (RSA) is also a critical issue in EON \cite{7105364}.
There are three main challenges related to RSA: (a) spectrum continuity- where the same set of subcarriers is to be used in every link throughout the lightpath similar to wavelength continuity requirement in WDM networks, (b) spectrum contiguity- where contiguous OFDM subcarriers are to be allocated for a demand
instead of using a full wavelength in a WDM network, and (c) spectrum non-overlapping- where the allocated spectrum is to be non-overlapping in nature. 

Currently research in the domain of EON largely focuses on two major aspects. In the first aspect, most of the existing studies on RSA  are based on the fixed shortest path routing algorithms and concentrate mainly on the spectrum allocation efficiency \cite{klinkowski2011routing, wang2011study}. Therefore, its major focus is on the network layer perspective of resource allocation. The second aspect mainly concentrates on impairment aware RSA in EON. Impairment aware RSA is again categorized into impairment mitigation \cite{ yang2012impairment, 7876823} and impairment aware cross-layer optimization \cite{zhao2015nonlinear, behera2016transmission}.

Transmission impairments,  such as shot noise, amplified spontaneous emission  (ASE) beat noise, crosstalk, nonlinear interference, and filter narrowing limit the transmission reach of the optical signals. There are two kinds of crosstalk: (a) In-band (or homodyne) crosstalk-  which is caused by factors like four-wave mixing, Rayleigh backscattering, and imperfect isolation from other signals transmitted using the same frequency\cite{tibuleac2010transmission,filer2012generalized}.  (b) Out-of-band (heterodyne) crosstalk- which is  caused by imperfect filter isolation from an adjacent channel. The  in-band crosstalk due to imperfect isolation is generated  when two or more optical signals co-propagate through the same cross-connect (XC)\cite{ramamurthy1999impact}.  In-band crosstalk is within the bandwidth of the primary signal and gets accumulated as the signal traverses through multiple XCs and multiple links of a particular lightpath.   Its impact   is significant and can become a major concern  when a particular lightpath consists of multiple hops.  The other important factor that can cause significant in-band crosstalk is the fiber nonlinearity as the network load increases.   Filter narrowing is another critical aspect to be considered in EON. Passband narrowing is caused due to limited bandwidth of Wavelength Selective Switch (WSS) compared to signal bandwidth and can be significant for long distance transmission  when the same signal traverses through multiple WSSs without optical-electrical-optical (O-E-O) regeneration\cite{tibuleac2010transmission}. Significant amount of works are available in the literature which addresses the impairment mitigation problem. The papers that deal with impairment mitigation, are the ones where physical layer impairments are considered while allocating spectrum to traffic demands. The major concerns of these papers are to avoid the detrimental effect of impairments while performing RSA. However, there are very few literature available for impairment aware cross-layer optimization in EON which is more significant and challenging. 

In that direction, if viewed from the perspective of optical transport layer, fragmentation is a critical issue that limits the effective use of network resources. Since EON allocates spectrum slots in contiguous manner, bandwidth fragmentation occurs as the size of contiguous slots increases. Even if adequate spectrum slots are available for a new connection,  it becomes impossible to accommodate  new requests, since spectrum slots are scattered. To overcome this serious issue, many approaches \cite{sone2011routing,talebi2014spectrum,zhang2014dynamic} are available in the literature. However, most of the studies address the fragmentation issue after the fragmentation has occurred in spectrum slots by means of connection rerouting. Connection rerouting increases traffic delay and  degrades performance by increasing system complexity. There are  few literatures available that consider the fragmentation as  part of the RSA \cite{wang2012spectrum,fadini2014subcarrier,patel2012routing}.  In \cite{patel2012routing}, authors have formulated the fragmentation problem as an Integer Linear Programming (ILP) model considering spectrum continuity and contiguity constraints. While,  Balanced Load Spectrum Allocation (BLSA) scheme presented in \cite{wang2011study} focuses on  load-balancing among network links from the beginning of every allocation which will have detrimental effects on delay-bandwidth product. Both of these papers have focused on the transport layer specific parametric optimization and have ignored the effect of physical layer impairments. 

On the other hand, in \cite{capucho2013ilp}, the objective function is mainly formulated to focus on the minimization of the network delay-bandwidth product. As a result, the  demands are scattered, creating serious fragmentation issues.  Minimizing delay-bandwidth product also means accommodation of  more and more demands on the shortest path, which in turn creates load imbalance. Having more users on a particular link can as well lead to high congestion which can further exhibit  non-linear effects. Moreover, survivability and restoration can become a serious concern which might necessitate  rerouting of all the connections of a congested link during link failure.    These observations prove that minimization of delay-bandwidth product or fragmentation have conflicting effects. Subsequently, the conditions will be even more severe if impairments are considered.

\begin{figure}
	\centering
	%	\hspace{-1.6cm}
	\includegraphics[trim={0cm 5cm 7cm 0cm}, height=.4\textwidth, width=.55\textwidth]{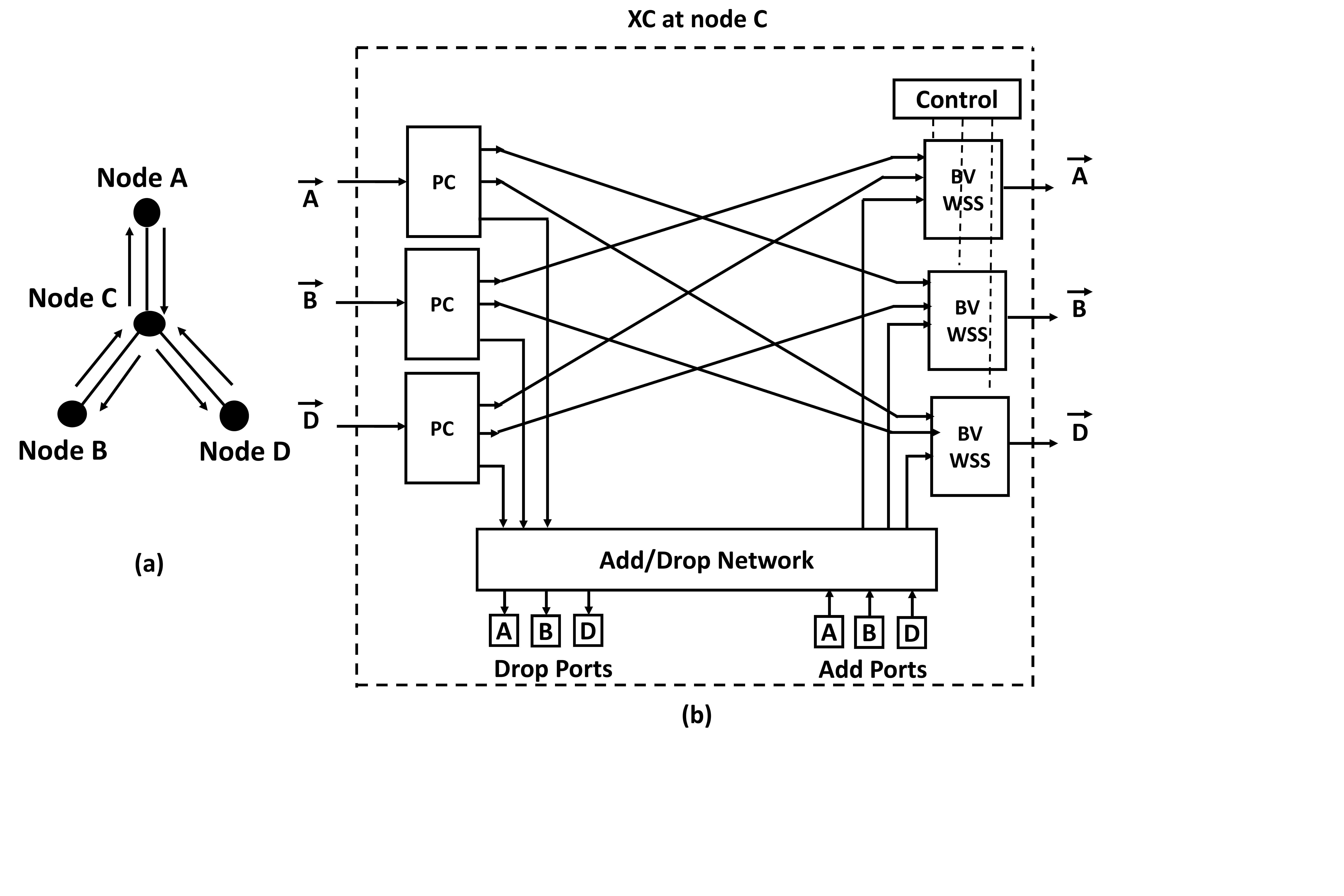}		
	\vspace{-1cm}
	\caption{Architecture of optical XC (3$\times$3 XC) for a CO-OFDM based EON: (a)A part of network; 4 nodes A,B,C,D namely are shown. (b)The architecture of optical XC at node C of the above considered part of the network.}
	\label{ArchXC}
\end{figure}

To address the delay-bandwidth product  along with fragmentation  and link congestion, we propose a novel cross-layer  joint optimization based RSA technique considering the physical layer impairments such as shot noise, ASE noise, in-band crosstalk, nonlinear interference and filter narrowing. The only other works that  considers such a cross-layer optimization framework is presented in \cite{zhao2015nonlinear, 7876823}. However, the authors of \cite{zhao2015nonlinear}  have considered only nonlinear impairments in cross-layer design for minimizing maximum frequency slot used on any link. This is similar to load balancing and therefore will have detrimental effects on delay.  Recently, in \cite{7876823}, the authors have considered only in-band crosstalk as impairments. They assume that in-band crosstalk might have detrimental effect, so they are minimizing the interactions among connections.  However, they have not calculated the actual interference and its effect on end-to-end link quality (SINR).
Our formulation  guarantees that a minimum end-to-end signal-to-interference-plus-noise ratio (SINR) for each link is maintained in presence of impairments. We  derive the probability of symbol error along with SINR for a CO-OFDM system employing 4-QAM.   Most of the time, joint optimization problems like ours are multi-objective in nature with associated proportionality constants which are often network dependent. However, in this paper, we propose joint optimization without considering the proportionality constants. 
In our previous work \cite{behera2016transmission}, we have addressed the minimization of delay-bandwidth product along with the transmission impairments. In this paper, first, we formulate the ILP problem for RSA as joint optimization in presence of impairments.  We next propose a heuristic algorithm with two ordering techniques, and compare them. To the best of our knowledge,  this is the first work  that  addresses the cross-layer joint optimization for RSA in  CO-OFDM based EON considering the physical layer impairments. 

The rest of the paper is organized as follows. In Section II the architecture for EON is described. The overview of crosstalk and its effect is presented in Section III. An ILP model and  novel heuristics are presented in Section IV and Section V respectively. Results and discussions are presented in Section VI and the concluding remarks are drawn in Section VII. Appendix A presents probability of symbol error derivation including noise and interference for 4-QAM.
\section{Network Architecture}
%\begin{figure}
%	\centering
%	%	\hspace{-1.6cm}
%	\includegraphics[trim={0cm 5cm 7cm 0cm}, height=.4\textwidth, width=.55\textwidth]{diagram.pdf}		
%	\vspace{-1cm}
%	\caption{Architecture of optical XC (3$\times$3 XC) for a CO-OFDM based EON: (a)A part of network; 4 nodes A,B,C,D namely are shown. (b)The architecture of optical XC at node C of the above considered part of the network.}
%	\label{ArchXC}
%\end{figure}

Various node architectures for EON has been presented in \cite{amaya2013introducing}. Fig. \ref{ArchXC} presents the schematic of the optical XC used for our study. A part of network  is shown in Fig. \ref{ArchXC}(a); A, B, C and D are the nodes. Architecture of optical XC at node C is shown in Fig. \ref{ArchXC}(b). The XC shown at node C is a 3$\times$3 XC where, the physical degree of node C is 3.  Based on the signal flow at XC, the channel may be pass-through channel, add-channel or drop-channel as in the WRNs. The first stage of the XC consists of  an array of passive combiners/splitters (PC)  and followed by a stage of bandwidth variable Wavelength Selective Switches (BV-WSS). Each BV-WSS is controlled by a controller which directs the appropriate subcarriers to their respective destinations.  The number of BV-WSS devices in an XC equals to the number of incoming links to the XC. The number of input ports of BV-WSS are equal to the links in that node. Each BV-WSS device has 1 output port which consists of $N$ subcarriers of the OFDM signal (total number of frequency slots (FSs)). From a node's perspective, each optical XC includes the facility for adding and dropping  channels, and  it must need separate add and drop ports corresponding to each nodes connected to the concerned XC node. For example, in Fig. \ref{ArchXC}(b), the optical XC at node C includes  separate add/drop ports for nodes A, B and D.

\section{Effect of impairments in EON}
In this section, we present an overview of impairments and their effect on SINR calculation.
%\vspace{-.2cm}
\begin{figure}[t!]
	\centering
	\begin{center}
		\hspace{-1cm}		
		\includegraphics[trim=0cm 13cm 0cm 0cm, height=.4\textwidth, width=.75\textwidth]{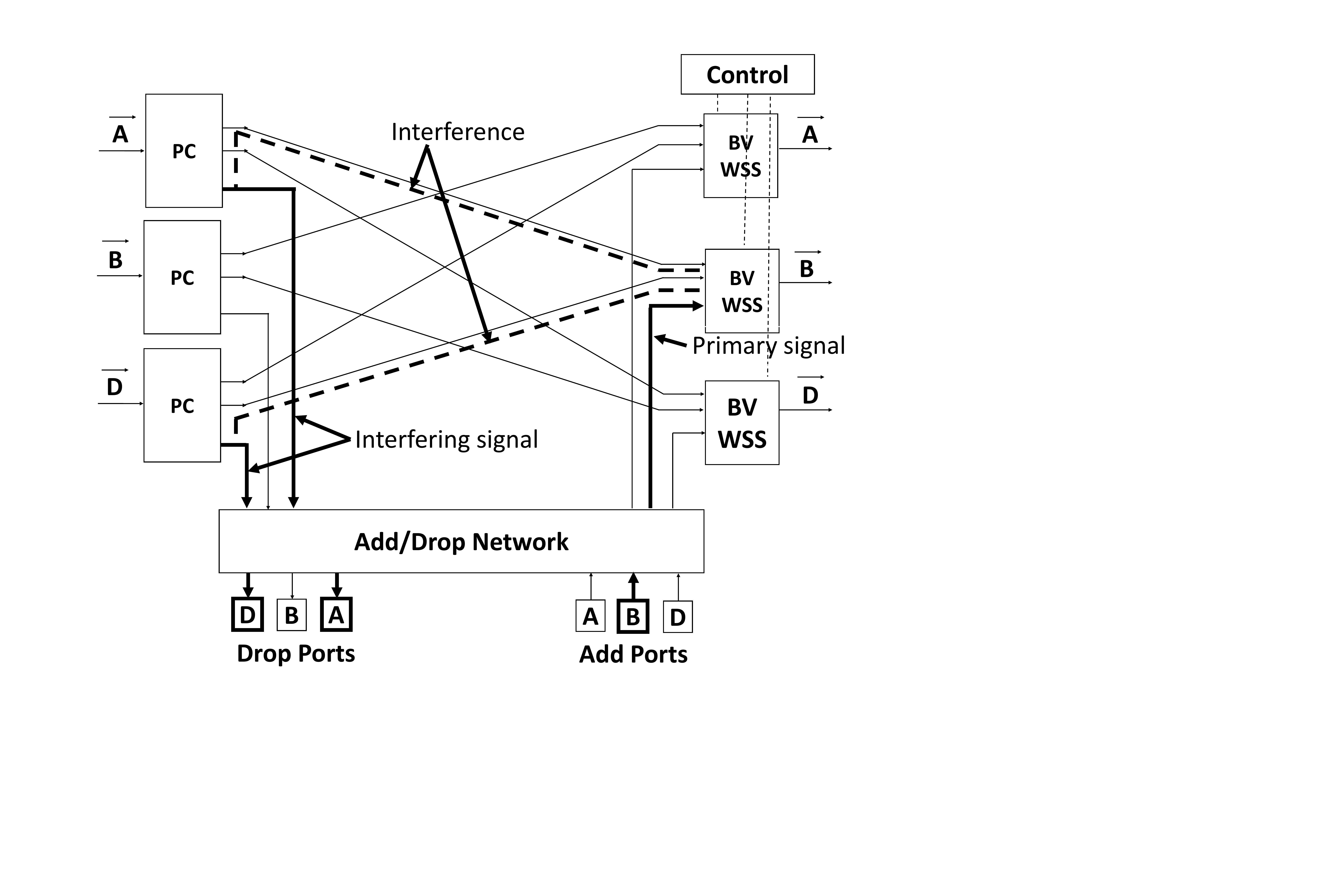}
		\label{adddrop}		
		\caption{Crosstalk in the add channel (XC is at node C): primary add channel and interfering drop channel }
		\label{pri_add}
	\end{center}
\end{figure}

\begin{figure}[!ht]
	\centering
	\includegraphics[trim=3cm 0cm 7cm 0cm, width=.5\textwidth, height=.4\textwidth]{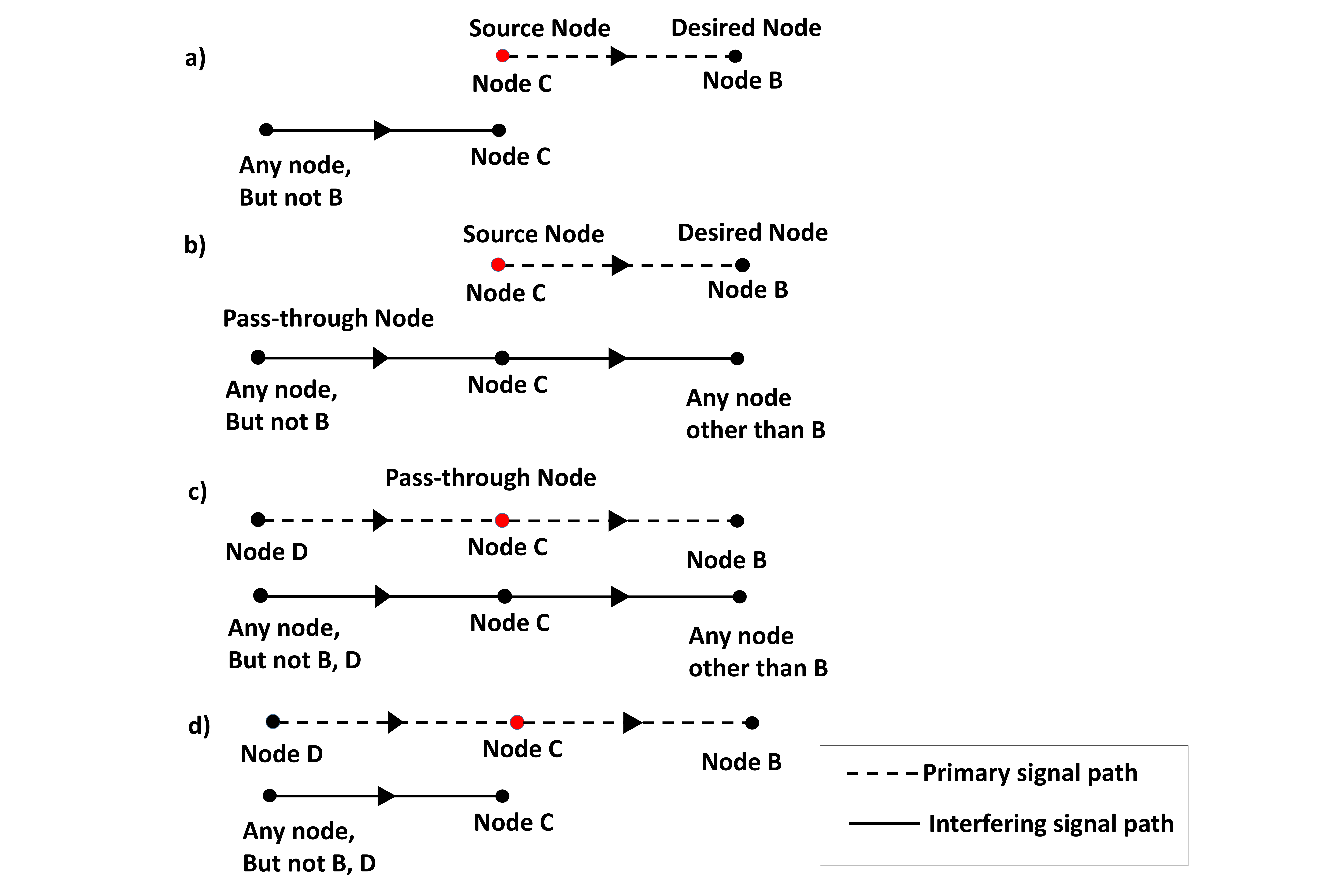}
	\caption{Cross-talk in an optical XC: a) primary add channel and interfering drop channel. b) primary add channel and interfering pass-through channel.  c) primary pass-through channel and interfering pass-through channel.  d) primary  pass-through channel and interfering drop channel}
	\label{crosstalkfigure}
\end{figure}

\subsection{Effect of crosstalk}
\subsubsection{Crosstalk in the add channel}
%\begin{figure*}[t!]
%	\centering
%	\begin{center}
%	\hspace{-1.5cm}	
%	\subfigure[]{\includegraphics[trim=0cm 12cm 7cm 0cm, height=.45\textwidth, width=.65\textwidth]{primary_add_interfere_drop}\label{adddrop}} \hspace{-3cm}
%	%		\captionsetup{justification=centering,margin=2cm}
%	\subfigure[]{\includegraphics[ trim=0cm 12cm 28cm 0cm, height=.45\textwidth, width=.5\textwidth]{primary_add_interfere_passthru}	\label{addpass}} 
%	\caption{Crosstalk in the add channel: (a) primary add channel and interfering drop channel (b) primary add channel and interfering pass-through channel.}
%\end{center}
%\end{figure*}
In this case, the primary signal is an add signal which is added at the XC, and goes to an outgoing node connected to the concerned XC. The interfering signal to this primary signal can be a drop channel signal or a pass-through channel signal if these signals satisfy certain direction of flow.

For example, let the primary signal be added at the XC targeting towards node B (refer Fig. \ref{pri_add}).

Condition to be satisfied by the other drop channel signal and pass-through signals to act as  interference to the considered primary signal:

\begin{itemize}
	\item  It can not come from node B as B to B connectivity is not there.
	\item Any other add or pass-through traffic to B is not a potential interferer as that can not be allocated using same frequency.
\end{itemize}
The interfering signals for the considered primary signal are shown in Fig. \ref{pri_add}. Since,
Fig. \ref{pri_add} depicts drop channels as the only source of interference, therefore, pass-through channel signals are not considered for interference (to A and D of pass-through nodes). Fig. \ref{crosstalkfigure} summarizes all four conditions for in-band crosstalk.
 Fig. \ref{crosstalkfigure}(a) and Fig. \ref{crosstalkfigure}(b) present the conditions of crosstalk in add channel where the source of interference are drop channels and pass-through channels respectively.

\subsubsection{Crosstalk in the pass-through channel}

The primary signal is a pass through channel signal, which is propagated through the node on which the considered XC  is attached, and goes to an outgoing node connected to the concerned XC. The interfering signal for this primary signal can be a drop channel signal or a pass-through channel signal if these signals satisfy certain direction of flow.

For example, let the primary signal be D to B through XC (refer Fig. \ref{crosstalkfigure}(c) and Fig. \ref{crosstalkfigure}(d)).

Condition to be satisfied by the other pass-through channel signals and drop channel signals to behave as  interference to the considered primary signal:

\begin{itemize}
	\item  It can not come from D and B as D and B are source-destination pairs.
	\item To B is not possible due to frequency clash.
	\item To A is not possible as A to A connectivity is not there.
\end{itemize}

Fig. \ref{crosstalkfigure}(c) depicts pass-through channels as the only source of interference, therefore, drop channel signals are not considered for interference. Similarly,  Fig. \ref{crosstalkfigure}(d) shows drop channel signals as the only source of interference and hence pass-through channels are not considered for interference.

%\textit{If the other signal  passes through the node, on which the considered XC  is attached (in our case, node C), from any node except the desired outgoing node of the XC for primary signal (in our case, node D to node B through node C), to  any other node} as depicted in Fig. 3(a). 
%%\ref{passpass}
%
%Condition to be satisfied by the other drop channel signals to behave as an interference to the primary signal:
%
%\textit{If the other signal is dropped from a link starting from a node other than the desired node of the primary signal (in our case, node D to node B through node C)   and ends in a node on which the considered  XC (in our case, node C)  is attached,} as shown in Fig. 3(b). 
%%\ref{passdrop}

The above described process can be applied to each node and associated optical XC in a particular lightpath, and  the effect of total crosstalk on the considered link can therefore  be calculated. 

\subsubsection{Crosstalk Calculation}
$P_{xtalk_l}$ for the $l$-th interferer can be calculated by multiplying the interference signal (same as primary signal $(P_r)$ in our case) with the crosstalk factor, $\epsilon_{{xtalk}}$. Hence, 
\begin{align}
P_{xtalk_l}={P_r} \epsilon_{{xtalk}} 
\label{xt}
\end{align} 

%\begin{figure*}
%	\centering
%	\begin{center}
%		%		\hspace{-1cm}		
%		\includegraphics[page=5, clip, trim=2cm 17cm 0cm 1.3cm]{Effect2.pdf}
%		%		\label{adddrop}		
%		%		\caption{Crosstalk in the add channel (XC is at node C): primary add channel and interfering drop channel }
%		%		\label{tab:table1}
%	\end{center}
%\end{figure*}

%	\begin{table*}[h!]
%		\centering
%		\caption{ILP PARAMETERS}
%%		\vspace{-.1in}
%		\label{param}
%		\begin{tabular}{c}
%			\includegraphics[page=5, clip, trim=2cm 17cm 0cm 2.5cm]{Effect2.pdf}
%		\end{tabular}
%	\end{table*}

%
%\begin{table*}[h!]
%	\label{param} 
%\centering
%
%\begin{tabular}{c}	
%	dfgrtytgr
%%\includegraphics[page=5, clip, trim=2cm 17cm 0cm 1.3cm]{Effect2.pdf}		
%\end{tabular}
%
%\end{table*}

\subsection{Effect of nonlinear interference}
We use the Gaussian noise model as described in  [Eq. (16),\cite{johannisson2014modeling}] for nonlinear impairments (NLI).

\begin{align} 
\label{NL}
&G_{span}^{NLI}(f)=\frac{3\gamma^{2}G(f)}{2\pi\alpha |{\beta_{2}}|}
\left[ G^{2}(f)\ln\left|\frac{\pi^{2}\beta_{2}(\triangle f)^{2}}{\alpha}\right| \right.\nonumber \\
&\left.+ \sum_{\substack{f'\\f'\neq f}} G^{2}(f)\ln\left(\frac{|f-f'|+\triangle f'/2}{|f-f'|-\triangle f'/2}\right) \right]
\end{align}

 where, $G(f)$ is signal PSD having signal bandwidth as $\triangle f$ and center frequency as $f$, $\beta$ is fiber dispersion, 
	$\gamma$ is fiber nonlinear coefficient and $\alpha$ is power attenuation, $\triangle f'$
	is the bandwidth of the other signal having center frequency as   $f'$. This approximation is valid for signal having bandwidth more than 28 GHz.

\subsection{Effect of filter narrowing}
FIlter narrowing in EON has been reported in 
	\cite{fabrega2016filter}. We use the simple approach described in \cite{heismann2010system} for reserving extra spectrum before spectrum allocation.

\subsection{SINR Calculation}

The probability of symbol error for the 4-QAM (all 4 symbols have same power) modulated subcarriers in the CO-OFDM signal is,

\footnotesize
\begin{align}
\label{proberror}
P_{e}= 2Q\left(\sqrt{SINR}\right)\left[1-\frac{1}{2}Q\left(\sqrt{SINR}\right)\right]
\end{align}
\normalsize
The derivation of $P_{e}$ along with SINR calculation is provided in the Appendix A.
 
% \vspace{-.2cm}
\section{ILP Formulation}
%	\begin{table*}[h!]
%		\centering
%		\caption{ILP PARAMETERS}
%		%		\vspace{-.1in}
%		\label{param}
%		\begin{tabular}{c}
%			\includegraphics[page=5, clip, trim=2cm 17cm 0cm 2.5cm]{Effect2.pdf}
%		\end{tabular}
%	\end{table*}
	\begin{table}[h!]
		\centering
		\caption{ILP PARAMETERS}
%		\vspace{-.1in}
		\label{param}
		\begin{tabularx}{\columnwidth}{cX}
			\toprule
			Symbol & Meaning\\
			
			\midrule
			
			\textbf{N}: & Total number of FS supported in each link\\ \\
			%			 \textbf{adjacency[i][j]}: & adjacency matrix; cost of the $\delta$ i-j in kilometers (Km)\\
			$\textbf{$\rho$}^\textbf{{s, d}}$: & Traffic matrix; number of frequency slots required for  \textit{(s-d)} pair\\ \\
			$\textbf{$\delta$}_\textbf{{i, j, r}}^\textbf{{s,d}}$:
			 & Binary; 1 if link i-j is present in the route 'r' of the \textit{(s-d)} pair, 	else 0\\	\\
			 	$\textbf{$\Delta_{i,j}$}$: & Link distance in Kms\\ \\
			 	$\textbf{d}_\textbf{{i,j}}$	: & Normalized link distance \\ \\
			 	$\textbf{W}_\textbf{i, j, k}$ : & Binary output; 1 if FS 'k' is the highest indexed FS used in link i-j, else 0\\ \\
			 	$\textbf{R}_\textbf{{i, j}}$ : & Integer; denotes the no of demands on link (i,j) \\ \\

			$\textbf{Pch}_\textbf{{r, k}}^\textbf{{s,d}}$: & Pre-calculated coherently received power in the route 'r' for the slot 'k' of the \textit{(s-d)} pair\\ \\
			
			$\textbf{SNR}_\textbf{{r,k}}^\textbf{{s,d}}$: & Pre-calculated SNR in the route 'r' for the slot 'k' of the \textit{(s-d)} pair\\ \\
			\textbf{NSIS}:& Constant; represents value of inverse of signal to interference plus noise(SINR) limit for non-selected routes\\ \\
			
			\textbf{SIS}: & Constant; represents value of inverse of SINR limit for selected route\\ \\
			$\textbf{L}_\textbf{r,f}^\textbf{s,d}$ : & Binary output; 1 if 'f' is the starting FS in the route 'r' of \textit{(s-d)} pair, else 0\\ \\
			
			$\textbf{X}_\textbf{i, j, k}$ : & Binary output; 1 if FS 'k' is used in link i-j, else 0\\ \\
			$\textbf{Pxt}_\textbf{r, f, k, i, j}^\textbf{s,d}$ :& Crosstalk power added in the XC  attached to the node 'i' of link $(i-j) \in r$ for the FS 'k' with the starting frequency slot 'f' for route 'r' for \textit{(s-d)} pair demand\\ \\
			
			$\textbf{P}_\textbf{r$^\prime$, f$^\prime$, k, j$^\prime$, i}^\textbf{s$^\prime$,d$^\prime$}$ : & Power of the interfering signal at node 'i' (calculated by \eqref{xt} \& \eqref{sinr2})  which propagates from the node j$^\prime$ to the cross-connect attached to node'i' for the FS 'k' with starting frequency 'f$^\prime$'\\ \\
			
			$\textbf{Pxt}_\textbf{r, k}^\textbf{s,d}$ : & Accumulated crosstalk power received in the FS 'k'  at the destination node 'd' for the \textit{(s-d)} pair\\ \\
			
				$\textbf{{Pnli}}_\textbf{{i,j,r,f}}^\textbf{{s,d}}$ : & Nonlinear interference power on frequency 'f' for \text{(s,d)} pair on route 'r' for link \text{$(i,j) \in r$}\\ \\
			$\textbf{{Pnli}}_\textbf{{r}}^\textbf{{s,d}}$ : & Nonlinear interference power  for \text{(s,d)} pair on route 'r'  \\
			
			\bottomrule
		\end{tabularx}
	\end{table}
In this section, we present our proposed ILP formulation. We contrast this ILP with respect to the  ILP for delay-bandwidth product optimization (termed as DBP ILP, henceforward) as given in \cite{capucho2013ilp}.  In the proposed multi-objective ILP, to minimize fragmentation along with delay-bandwidth product and congestion, we propose a new modified optimization framework.  
The SNR for each FS  in all the routes of each \textit{(s-d)} pair is  pre-calculated by  \eqref{pch} and \eqref{sinr1}. The bandwidth of a connection is calculated by multiplying slot width  with number of slots requested. For filter narrowing, extra reserved spectrum is pre-calculated and adjusted into $\rho^{s,d}$ depending upon the route selected. The ILP parameters are presented in Table \ref{param}.

\subsection{Proposed ILP formulation}
Our objective is different from that of \cite{capucho2013ilp}  to account for fragmentation, congestion and delay-bandwidth product in a single function. As mentioned before, we propose multi-objective optimization without considering proportionality constants. \\
\textit  {minimize:} 
\begin{align}
\underset{i}{\sum}\underset{j}{\sum}\underset{k}{\sum}
 \frac{R_{i,j} 
 	\times d_{i,j}} {N- {k 	\times 	W_{i,j,k}}} 
 \label{obj}
\end{align} 
 Knowing that $W_{i,j,k}$ indicates whether $k$ is the highest indexed slot used in link (i-j) (if it is 1 for a particular $k$),  the denominator in the objective function denotes the contiguous free  slots available in  a link after the highest occupied slot. As we are minimizing inverse of that, the process ensures that minimum fragmentation is happening in each link. The numerator when summed over all links denotes the delay-bandwidth product.  Therefore, the ratio ensures that the congestion is avoided in each link.\\
\textit {Subject to:} \\
Single path routing constraint:
\begin{equation}
\label{constraint1}
\begin{aligned}
\sum_{r}  \sum_{f} \textbf{$L_{r,f}^{s,d}$} = 1\; ; \:\: 
\begin{cases}
&\forall (s,d) \;\text{pairs} ,\\
& \text{where},  \; r\in R \text{ on each} \;(s,d) \; \text{pair},\\
& f=\{1,2, \ldots, \text{N-$\rho^{s,d}$} \}
\end{cases} 
\end{aligned}
\end{equation}
Constraints (\ref{constraint1}) refers to the single path routing constraint 
	which ensures that only one route and only one 'first frequency slot' for the selected route for each \textit{(s,d)} pair is selected.\\
Spectrum  contiguity constraints:
\begin{equation}
\label{constraint2}
\begin{aligned}
\textbf{$L_{r,f}^{s,d}$} \leq \textbf{$X_{i,j,k}$}\; ; \:\: 
\begin{cases}
&\forall (s,d) \;\text{pairs} ,\\
&\forall \text{ link i-j in} \text{ route r of } (s,d)\text{ pair}, \\
& \text{where},  \; r\in R \text{ on each} \;(s,d) \; \text{pair},\\
& f=\{1,2, \ldots, \text{N-$\rho^{s,d}$} \},\\
& k=\left\lbrace f,f+1, \dots, f+\text{$\rho^{s,d}$} \right\rbrace 
\end{cases} 
\end{aligned}
\end{equation}
Constraints \eqref{constraint2} ensures that first slot, for each demand,
cannot exceed $N-\rho^{s,d}$. 

Spectrum continuity and non-overlapping constraint:
\begin{equation}
\label{constraint3}
\hspace{-2.5cm} \begin{aligned}
\sum_{s} \sum_{d} \sum_{r} \sum_{f} \text{$\delta_{i,j,r}^{s,d}$}\times\textbf{$L_{r,f}^{s,d}$} \;\leq\; \textbf{$X_{i,j,k}$}\; ; \:\: \\ 
\begin{cases}
&\forall  k=\left\lbrace 1,2, \dots, N \right\rbrace, \\
&\forall \text{ link i-j } \text{ pair}, \\
& \text{where},  \; r\in R \text{ on each} \;(s,d) \; \text{pair},\\
& f\;\leq \; k\;,f \; \geq \; k- \text{$\rho^{s,d}$} \; ,\\
& f \; \leq \; N- \text{$\rho^{s,d}$} \;, \;f\;\geq \; 1
\end{cases} 
\end{aligned}
\end{equation}
 Constraints \eqref{constraint3} attribute to the allocation of same FSs in every link along the selected route of each \textit{(s,d)} pair.\\

Link capacity constraint:

\begin{equation}
\label{constraint4}
\begin{aligned}
\sum_{k}  \textbf{$X_{i,j,k}$} \leq N\; ; \:\: 
\forall (i,j) \;\text{link pairs in physical network} .\\ 
\end{aligned}
\end{equation}

\noindent Constraints (\ref{constraint4}) ensures that the number of FSs allocated in each link in the physical topology is less than or equal to the link capacity ($N$).

Fragmentation and delay constraints are given below:
 \begin{equation}
 \label{constraint5}
 \begin{aligned}
 R_{i,j}=\underset{s}{\sum}\underset{d}{\sum}
 \underset{r}{\sum}
 \underset{f}{\sum}
 L_{r,f}^{s,d}
 \times
 \delta_{i,j,r}^{s,d} 
 \begin{cases}
 \forall (i,j) \text { link pairs}, \\ 
f=\{1,2, \ldots, \text{N-$\rho^{s,d}$} \}\\
 \end{cases}
 \end{aligned}
 \end{equation}  
 \begin{equation}
 \label{constraint6}
 \begin{aligned}
  W_{i,j,k'} & \leq 1- \frac{\sum_{k=k'+1}^{N} X_{i,j,k}} {N} \; ; \:\:
  \begin{cases}
  \forall (i,j) \text { link pairs}, \\ 
     k' =\left\lbrace 1,2, \dots, N-1 \right\rbrace \\
  \end{cases}
 \end{aligned}
 \end{equation} 
  \begin{equation}
  \label{constraint7}
  \begin{aligned}
 \hspace{-2.3cm} W_{i,j,k'} & \leq  X_{i,j,k'} \; ; \:\:
  \begin{cases}
  \forall (i,j) \text { link pairs},  \\
   k'=\left\lbrace 1,2, \dots, N \right\rbrace \\
  \end{cases}
  \end{aligned}
  \end{equation} 
%  \vspace{-.4cm}
  \begin{equation}
  \label{constraint8}
  \begin{aligned}
\hspace{-1.3cm} \sum_{k'} W_{i,j,k'} & \geq \frac {\sum_{k'} X_{i,j,k'}} {N} \; ; \:\:
  \begin{cases}
  \forall (i,j) \text { link pairs}  \\  
  \end{cases}
  \end{aligned}
  \end{equation}
  Constraints \eqref{constraint5} denote the total number of demands that are present on link (i,j).
  Constraints (\ref{constraint6}-\ref{constraint8}) are responsible for selecting the highest indexed slot used on link (i,j) for any demand.

% \vspace{-1.4cm}
\subsection{Effect of Impairments}
%\vspace{-.5cm}
 In this subsection, we show that how we have incorporated the effect of impairments.   \\
 Crosstalk contraints:
 \begin{equation}
 \label{constraint9}
 \begin{aligned}
 &\text{Pxt}_\text{r,f,k,i,j}^\text{s,d} =
 \sum_{s^\prime} \sum_{d^\prime}\sum_{r^\prime}\sum_{f^\prime} \sum_{j'} \sum_{i}\text{P}_\text{r$^\prime$,f$^\prime$,k, j$^\prime$,i}^\text{s$^\prime$,d$^\prime$}   \times\text{$L_{r^\prime,f^\prime}^{s^\prime,d^\prime}$}\; ; \:\: \\
 &\hspace{1.5cm}\begin{cases}
 &\forall  \; (s,d) \; \text{pair} , (s,d) \neq (s^\prime,d^\prime),\\
 &\forall (i,j)\in r ,\\
 &\forall r\in R \text{ on each} \;(s,d) \; \text{pair},\\
 &\forall f=\{1,2, \ldots, \text{N-$\rho^{s,d}$} \},\\
 &\forall k=\left\lbrace f,f+1, \dots, f+\text{$\rho^{s,d}$} \right\rbrace ,\\ 
% & (s,d) \neq (s^\prime,d^\prime), \\
 &\text{where},  \; r^\prime\in R \text{ on each} \;(s^\prime,d^\prime) \; \text{pair},\\
 &f^\prime\;\leq \; k\;,f^\prime \; \geq \; k- \text{$\rho^{s^\prime,d^\prime}$} \; ,\\
 &f^\prime \; \leq \; N- \text{$\rho^{s^\prime,d^\prime}$} \;, \;f^\prime\;\geq \; 1, \\
 \end{cases} 
 \end{aligned}
 \end{equation}
 \vspace{-.2cm}
 \begin{equation}
 \label{constraint10}
 \begin{aligned}
 &\hspace{-1cm}\text{Pxt}_\text{r,k}^\text{s,d} =\;  \sum_{i} \sum_{j} \sum_{f}\text{P{xt}}_\text{r,f,k,i,j}^\text{s,d}\; ; \:\: \\
 &\hspace{.7cm}\begin{cases}
 &\forall  \; (s,d) \; \text{pair}, \\
 & \forall  \; r\in R \text{ on each} \;(s,d) \; \text{pair},\\ 
 & \forall k=\left\lbrace 1,2, \dots, N \right\rbrace ,\\
 &\text{where},  \;f\leq \; k\;,f \geq \; k- \text{$\rho^{s,d}$} \; ,\\
 &f \leq \; N- \text{$\rho^{s,d}$} \;, \;f\geq \; 1, \\
 \end{cases} 
 \end{aligned}
 \end{equation}
 
% \begin{equation}
% \label{constraint11}
% \begin{aligned}
% &\hspace{0cm}\frac{\text{$Pxt_{r,f,k}^{s,d}$}} {\text{$P{ch}_{r,k}^{s,d}$}}  +  \frac{{L}_{r,f}^{\text{s,d}}}{\text{$SNR_{r,k}^{s,d}$}} + NSIS\times{L}_{r,f}^{s,d}  <=  NSIS \; + SIS\; ;  \\
% &\hspace{1.7cm}\begin{cases}
% &\forall  k=\left\lbrace 1,2, \dots, N \right\rbrace, \\
% &\forall \; \textit{(s-d)} \text{ pair}, \\
% & \forall r\in R \text{ on each} \;(s,d) \; \text{pair},\\
% &\text{where},\;f\;\leq \; k\;,f \; \geq \; k- \text{$\rho^{s,d}$} \; ,\\
% &f \; \leq \; N- \text{$\rho^{s,d}$} \;, \;f\;\geq \; 1 
% \end{cases} 
% \end{aligned}
% \end{equation}
 \noindent Constraints (\ref{constraint9}) and (\ref{constraint10}) are meant for crosstalk calculation. For each link 'i-j' of route r of  each \textit{(s,d)} pair, the interfering power from all other \textit{(s',d')} pairs are enumerated in constraint (\ref{constraint9}). Summing over  all the above mentioned power for each link in each \textit{(s,d)} pair till the destination node gives the total crosstalk power received at the destination node as given by constraints (\ref{constraint10}).
 
 Nonlinear impairment constraints:
 
  \begin{equation}
  \label{NLI1}
  \begin{aligned}
  &\hspace{0cm}\text{Pnli}_\text{i,j,r,f}^\text{s,d} =\; \:\:\\
  & {\Omega \triangle {f_m} G(f+\frac{\rho^{s,d}}{2})}\left[ G^{2}(f+\frac{\rho^{s,d}}{2})\ln\left|\frac{\pi^{2}\beta_{2}(\triangle f)^{2}}{\alpha}\right| \right. \\
  &\left.+\sum_{s'}\sum_{d'} \sum_{r'} \sum_{f'} G^{2}(f'+\frac{\rho^{s',d'}}{2})\ln(\mu) \right] \times\text{$L_{r^\prime,f^\prime}^{s^\prime,d^\prime}$} \times \delta_{i,j,r'}^{s',d'}\; ; \:\: \\ 
  & where,
  \Omega=\frac{3 \gamma^{2}}{2\pi\alpha |{\beta_{2}}|}; 
   \mu=\frac{f+\frac{\rho^{s,d}}{2}-f'-\frac{\rho^{s',d'}}{2}+\frac{\triangle {f}'}{2}}{f+\frac{\rho^{s,d}}{2}-f'-\frac{\rho^{s',d'}}{2}-\frac{\triangle {f}'}{2}} \\
  &\hspace{1cm}\begin{cases}
  &\forall  \; (s,d) \; \text{pair}, (s,d) \neq (s^\prime,d^\prime),\\
  &\forall r\in R \text{ on each} \;(s,d) \; \text{pair},\\ 
  &\forall f=\{1,2, \ldots, \text{N-$\rho^{s,d}$} \},\\
   &\forall \; \text{ link $(i,j) \in r $ } \;  \\
  &\text{where},  \; f'=\{1,2, \ldots, \text{N-$\rho^{s',d'}$} \},\\
    & r^\prime\in R \text{ on each} \;(s^\prime,d^\prime) \; \text{pair},\\
 
  \end{cases} 
  \end{aligned}
  \end{equation}
  
   \begin{equation}
   \label{NLI2}
   \begin{aligned}
   &\hspace{0cm}\text{Pnli}_\text{r}^\text{s,d} =\;  \sum_{i} \sum_{j} \sum_{f} \text{P{nli}}_\text{i,j,r,f}^\text{s,d} \times \text{$L_{r,f}^{s,d}$} \times \ceil[\Bigg]{\frac{\Delta_{i,j}}{\text{span length}}} \; ; \:\: \\
   &\hspace{1cm}\begin{cases}
   &\forall  \; (s,d) \; \text{pair}, \\
   & \forall  \; r\in R \text{ on each} \;(s,d) \; \text{pair},\\   
   & \text{where},  \; f=\{1,2, \ldots, \text{N-$\rho^{s,d}$} \},\\     
   \end{cases} 
   \end{aligned}
   \end{equation}

\noindent  Constraints \eqref{NLI1} and \eqref{NLI2}  attribute to nonlinear interference. Constraints \eqref{NLI1} meant for each span of $\text{(i-j)}$ link of a particular  $\text{(s,d)}$ pair (having center frequency $f+\rho^{s,d}/2$) where interference is calculated from all other $\text{(s',d')}$ pairs existing on same link $\text{(i-j)}$ (having center frequency $f'+\rho^{s',d'}/2$). Constraints \eqref{NLI2} calculates NLI for a particular $\text{(s,d)}$ pair for each route. 
 
 \begin{equation}
 	\label{constraint11}
 	\begin{aligned}
 		&\hspace{-1cm}\frac{\text{Pnli}_\text{r}^\text{s,d}} {\text{Pch}_\text{r,k}^\text{s,d}} + \frac{\text{Pxt}_\text{r,k}^\text{s,d}} {\text{Pch}_\text{r,k}^\text{s,d}}  +  \frac{{L}_{r,f}^{\text{s,d}}}{\text{SNR}_\text{r,k}^\text{s,d}} + NSIS\times{L}_{r,f}^{s,d} \; \\  		
 		& \hspace{3cm}\leq  NSIS \; + SIS\; ;  \\
 		&\hspace{-1cm}\begin{cases}
 			& k=\left\lbrace 1,2, \dots, N \right\rbrace, \\
 			&\forall \; \textit{(s-d)} \text{ pair}, \\
 			& \forall r\in R \text{ on each} \;(s,d) \; \text{pair},\\
 			&\text{where},\;f\;\leq \; k\;,f \; \geq \; k- \text{$\rho^{s,d}$} \; ,\\
 			&f \; \leq \; N- \text{$\rho^{s,d}$} \;, \;f\;\geq \; 1 
 		\end{cases} 
 	\end{aligned}
 \end{equation}
 
\noindent Constraints (\ref{constraint11}) are the transmission impairment constraints for the minimum SINR requirement corresponding to maximum allowable BER. For high local oscillator (LO) power, this SINR constraints account for the  LO-ASE beat noise, LO-NLI and LO-crosstalk beat noise. A set of FSs are alloted to a  demand if and only if there exists a path on which the SINR value is greater than the minimum allowable SINR ($SINR_{th}$) for a lightpath. Two constant parameters $NSIS$ and $SIS$ are added  to form a linear constraint from a non-linear constraint. That means, for selected paths, the value of inverse of SINR should be less than or equal to $SIS$ since ${L}_{r,f}^{s,d }$ = 1 for the selected paths and for the non-selected paths, the value of inverse of SINR should be less than a high value. Here this value is limited by $NSIS + SIS$ since ${L}_{r,f}^{s,d}$ = 0. Therefore, it is noted that $NSIS \gg SIS$.
%\vspace{-.4cm}

\subsection{Linearization} 
Here, we present the linearization of all equations. The objective function \eqref{obj} can be simplified as follows,
Let 
\begin{align}
& \frac{R_{i,j}}{N-k \times W_{i,j,k}}=\psi_{i,j,k} \label{first eq}\\
& \Rightarrow R_{i,j}=N \times \psi_{i,j,k}- k \times W_{i,j,k} \times \psi_{i,j,k} \label{last eq} \\
& \Rightarrow R_{i,j}=N \times \psi_{i,j,k}- k \times Z_{i,j,k} 
\end{align}
Eq. \eqref{last eq} contains two variables  $W_{i,j,k}$ (binary) and $\psi_{i,j,k}$ (continuous) whose multiplication gives rise to non-linearity. The solver we are using is capable of solving only linear equations. So for linearization, we define a new variable $Z_{i,j,k}$ which is given by, 
\begin{align}
\label{nonlinear}
&Z_{i,j,k}=W_{i,j,k} \times \psi_{i,j,k} \nonumber\\    
&Z_{i,j,k} \geq 0 \nonumber\\
&Z_{i,j,k} \leq \bar{\psi} W_{i,j,k} \nonumber\\
&Z_{i,j,k} \leq \psi_{i,j,k} \nonumber\\
&Z_{i,j,k} \geq \psi_{i,j,k}-\bar{\psi} (1-W_{i,j,k})
\end{align}
where $\bar{\psi}$ is the upper bound for $\psi_{i,j,k}$.

Now we can modify the new objective function as,\\
\textit {minimize:} 
\begin{equation}
\label{mod_obj}
\underset{i}{\sum}\underset{j}{\sum}\underset{k}{\sum} \psi_{i,j,k}
\times d_{i,j} 
\end{equation}
subject to all the constraints (\ref{constraint1}-\ref{nonlinear}) mentioned above. Similarly, \eqref{NLI2} contains $\text{Pnli}_\text{i,j,r,f}^\text{s,d}$ (continuous) and  $L_{r,f}^{s,d}$ (binary) variable which can be linearized similar to \eqref{nonlinear}.

\section{Heuristic Algorithms}
The RSA problem without impairments was proven to be a NP-Hard problem \cite{wang2011study}. The ILP formulation can be evaluated only for small networks. Therefore, for larger networks we propose near optimal heuristics. In this section, we  propose a novel heuristic based on two different demand ordering techniques. Thereafter, the complexity analysis of the proposed heuristic is also presented.
%\vspace{-.3cm}
\subsection{Ordering of demands}
Optimality of a static RSA heuristic algorithm depends upon the order in which the demands are processed. Considering our objective function, we propose a novel ordering technique, Most Congested Demands First (MCDF)  and compare with a existing ordering technique,  Most Subcarriers First (MSF) \cite{7105364}. 
\begin{itemize}
	\item [1.] Most Subcarriers First (MSF): Here, the demand which has requested highest number of FSs are processed first. 
	\item[2.] Most Congested Demands First (MCDF): 
	Let $D$ denotes the set of demands and $P_{d_e}$ denotes the $K$-shortest paths associated with demand $d_e \in D$. Let $l_p$ denotes link on path $p^k \in P_{d_e}$, $k \in \left\lbrace1,2, \ldots, K \right\rbrace$. Let $\rho^{d_e}$ denotes the traffic of demand $d_e$. Let $P$ denotes set of all paths \textit{i.e}, $P=\underset{{d_{e}} \in D}{\bigcup}P_{d_{e}}$. Let $P_{l_p}$ denotes set of paths that go through link $l_p$ such that $P_{l_p} \subset P$. We define a binary variable $I_{d_e}$ which is set to 1 if $P_{d_{e}}\cap P_{l_{p}}\neq\phi$ else set to 0. Now for each link $l_p$ we define a congestion metric $C_{l_p}=\underset{{{d_{e}}\in D}}\sum I_{d_e} \times \rho^{d_e}$. The length of each path can be given as
	$L_{p^k}=\underset{{l_p} \in p^k} {\sum} C_{l_p}$. 
	Note that, for each demand, $L_{p^k}$ will have $K$ values corresponding to $K$-shortest paths, $1^{st}$ value referring to $k=1$, $2^{nd}$ value referring to $k=2$ and so on.  Now for each demand we calculate $G_k=L_{p^k} \times (1-\frac{delay(p^k)}{delay_{max}})$, where $delay({p^k})$ is the delay associated with path $p^k$ of demand $d_e$ and $delay_{max}$ is the maximum delay for any route in the network. The first term in the multiplication refers to congestion whereas the second term favors  shortest path as it is weighted by a large number \textit{i.e} $delay_{max}$.   Let $G_{d_{e}}=\frac{\underset{k}{\sum}G_{k}}{K}$   denotes the average value of $G_k$ for demand $d_e$. This is repeated for all the demands ${d_e} \in D$. Then  demands are processed in descending order of $G_{d_e}$. 
	The motivation behind MCDF ordering is to schedule the highest congested demands first.  MCDF takes into consideration of all the candidate paths of all demands before selecting a particular demand for scheduling unlike  MSF ordering. 
\end{itemize}

%\begin{figure}
%	\centering
%	\includegraphics[trim=0cm 3cm 2cm 4cm, height=.25\textwidth, width=.4\textwidth]{Llp1}
%	\caption{MCDF ordering}
%
%	\label{BSF}
%\end{figure}

\subsection{Proposed Heuristic Algorithm}
In Algorithm 1, we provide a pseudo code for our heuristic.
We serve the connections sequentially with the orderings as described above. Let  $SA(d_e)$ denotes spectrum allocation function and returns the best possible accessible path for demand $d_e$. $SA(d_e)$ function takes into consideration of all $K$-path of a selected demand and calculate $\psi_{i,j,k} \times \Delta_{i,j}$ according to \eqref{first eq}. The evaluated parameter is then summed over all links of the selected path. The path which yields the minimum value of $ \underset{{(i,j)} \in p^k} {\sum} \psi_{i,j,k} \times \Delta_{i,j} $ is chosen for spectrum allocation   (refer mark 2 of Algorithm 1). We begin our allocation by using First-Fit algorithm \cite{jinno2010distance} from the output  returned by the $SA(d_e)$ function.   
Now, we check all the possible slot positions available for satisfying  $\rho^{d_e}$ on the selected path based on previous allocations and store them in matrix $FF()$ (refer mark 4 of Algorithm 1).
For each link $l_p$ of the selected path, we calculate the interference for the first accessible slot stored in $FF()$ up to its $\rho^{d_e}$,  from existing connections according to the conditions given in Section III (refer mark 6 of Algorithm 1). The interference on the same slot (and hence on the same frequency) for all links of the same path can simply be added to give aggregate interference.    Then we calculate the $SINR$ using  (\ref{pch}-\ref{sinr2}) and this $SINR$ is compared with minimum $SINR$ required for establishing a connection i.e $SINR_{th}$. If $SINR \geq SINR_{th}$, then we allocate the spectrum  else we check for next slot available up to its $\rho^{d_e}$ in  $FF()$ on that path and repeat the same conditions above, till we get free slots for satisfying the requested demand in consideration (refer mark 8, 9 and 10 of Algorithm 1).
The algorithm tries to find the requested slots on the selected path, taking into consideration of all the previously allocated demands and  the required spectrum continuity and  contiguity constraints. If no suitable slots are found on the selected path, then the process is repeated for the next best path returned by $SA(d_e)$ function and so on. A connection is blocked if all paths returned by $SA(d_e)$ are not suitable for spectrum allocation (refer mark 15 of Algorithm 1).  Once a demand is allocated, the slots of the associated links on the selected path are blocked for all other connections which share the same links.

\begin{algorithm}
	\caption{Heuristic Algorithm}
	\begin{algorithmic}[1]		
		\State Sort the demands according to one of the ordering policies mentioned above ($D_s$).
		\For{each demand $i \in D_s$ (in order)}	
		
		Invoke $SA(d_e)$ to find the best possible path 
		\For{each selected path} (from best possible path 
		
		\hspace{.5cm} onwards)
		
		\State $FF () \leftarrow $ First-Fit algorithm (to find all possible
		
		\hspace{.9cm}  slot positions)

		\For{$m=FF(j):FF(j)+\rho^{d_e} (j=1,2,...)$}
		
		\For{each link $l_p$ of the selected path}	
		
		\hspace{0.7cm} Calculate interference for the possible slots
		
		\hspace{.7cm} from 	existing connections according to the
		
		\hspace{.8cm} conditions 		given in Section III
		\EndFor	
		
		\State Calculate aggregate interference for all links of 
		
		\hspace{.7cm}	the  selected path
		
		\State  Calculate $SINR$
		
		\If{$SINR \geq SINR_{th}$}

		\hspace{.6cm} Assign the requested spectrum from $FF(j)$ to
		
		\hspace{.5cm} $FF(j)+\rho^{d_e}$ 		and block those slots in all other
		
		\hspace{.5cm} links shared by  other connections.
		
		\Else \hspace{.2cm} {(if $SINR < SINR_{th}$)}
		
		\hspace{.7cm}	$j=j+1$
		\EndIf

		\EndFor
		\EndFor
		\State Block the connection if no path is suitable for spec-
		
		trum allocation.
		\EndFor
		
	\end{algorithmic}
\end{algorithm}

\subsection{Complexity Analysis and Convergence of Heuristics}
The computational complexity of our proposed heuristics is bounded by $O(|D^2||P^3||N^2||L|)$, where $|D|$, $|P|$ and $|L|$ are  total number of demands, paths and links respectively. Since we are tearing down connections in each step of the algorithm, the heuristic is guaranteed to converge in finite steps.

\section{Results and Discussion}
%\vspace{-.1cm}
\begin{table}[]
	\centering
	\caption{SIMULATION PARAMETERS}
	\vspace{-.1in}
	\label{sim param}
	\begin{tabular}{|l|l|}
		\hline
		Parameter                                  & Value     \\ \hline
		Local Oscilator power ($P_{lo}$), Received power ($P_r$)                & 0 dBm, -12 dBm     \\ \hline
		%		Received power ($P_r$)                        & -12 dBm   \\ \hline
		Responsivity ($R_a$), Operating Wavelength                            & 0.7 A/W, 1550 nm      \\ \hline
		%		Operating Wavelength                       & 1550 nm   \\ \hline
		Spontaneous Emission factor ($n_{sp}$)          & 2         \\ \hline
		EDFA gain ($G$), EDFA spacing and span length                             & 21 dB, 100 Km     \\ \hline
		Fiber attenuation ($\alpha$), WSS loss                          & 0.2 dB/Km, 2 dB \\ \hline
		%		WSS loss                                    & 2 dB \\ \hline
		%		EDFA spacing                               & 100 Km    \\ \hline
		Crosstalk factor ($\epsilon_{xtalk}$)        & -18.5 dB  \\ \hline
		Nonlinear coefficient ($\gamma$)  &  1.33 \text{$W^{-1} Km^{-1}$}\\ 
		\hline
		fiber dispersion ($\beta_2$) & -21.7 \text{$ps^2/Km$} \\
		\hline
		Inverse SINR for selected route ($SIS$)      & 1/32      \\ \hline
		Inverse SINR for non-selected route ($NSIS$) & 200       \\ \hline
		SINR threshold ($SINR_{th}$)								& 15 dB at  \\
		& $P_{e}=10^{-9}$ \\ \hline	
		
	\end{tabular}
\end{table}
In this section, we compare our proposed ILP with the DBP ILP, the heuristics and  with other existing scheme. The simulation parameters can be found in Table \ref{sim param}. 
Demand requests between (\textit{s-d}) pairs are predetermined. For each (\textit{s-d}) pair, $K=3$  paths are preset.   The algorithm results are averaged over 20 different set of similar demands.  We used IBM CPLEX Optimization Studio \cite{ibm2014ibm} for ILP and MATLAB for  algorithms.  We have evaluated the performance for a small network as well  for a realistic network.

\begin{figure*}[t!]
	%	\centering
	\begin{center}
%		\vspace{-0.5cm}
		\hspace{-1cm}	
		\subfigure[]{\includegraphics[height=.3\textwidth, width=.33\textwidth]{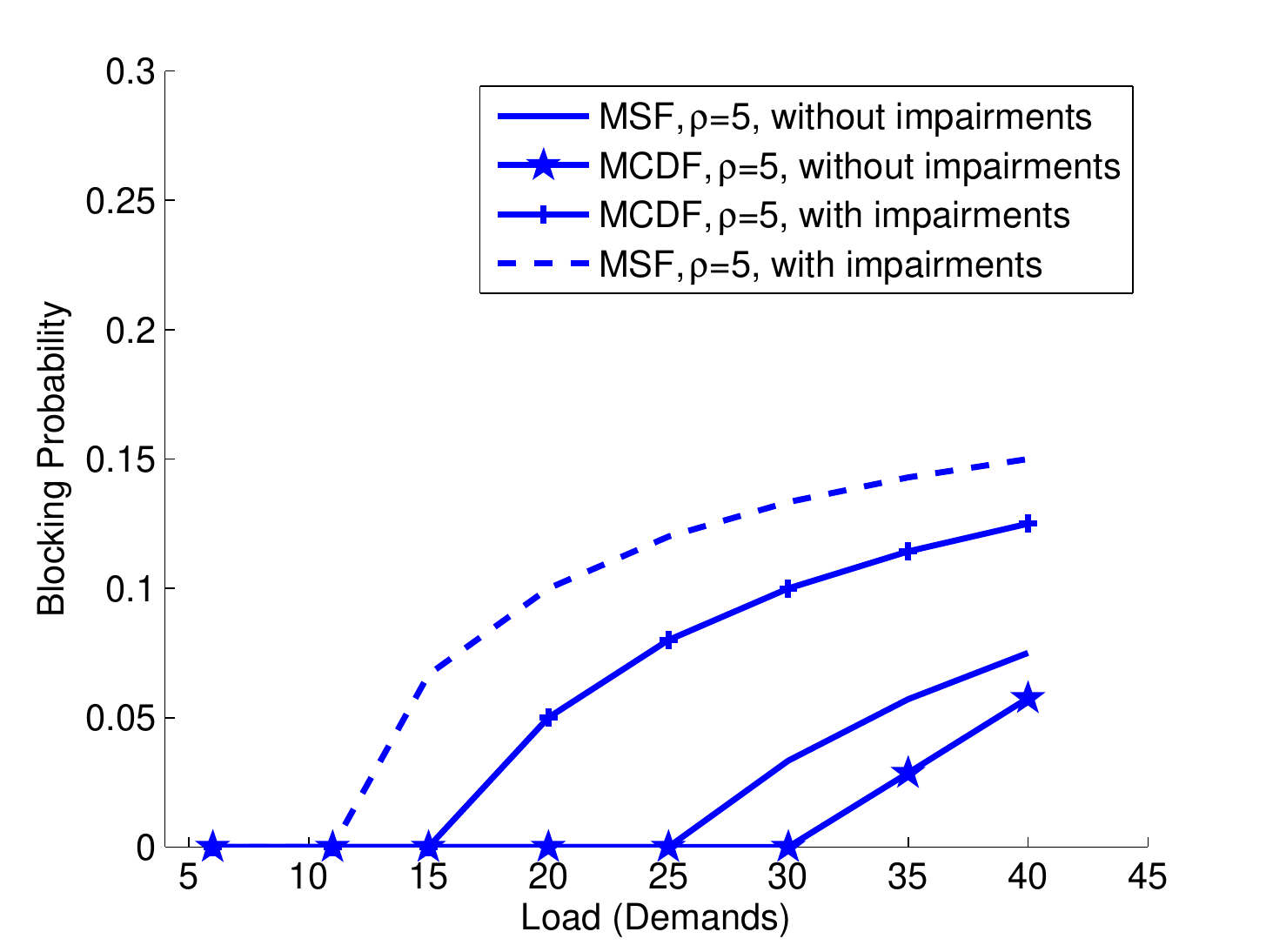}\label{load_bbp}} 
		\subfigure[]{\includegraphics[height=.3\textwidth, width=.33\textwidth]{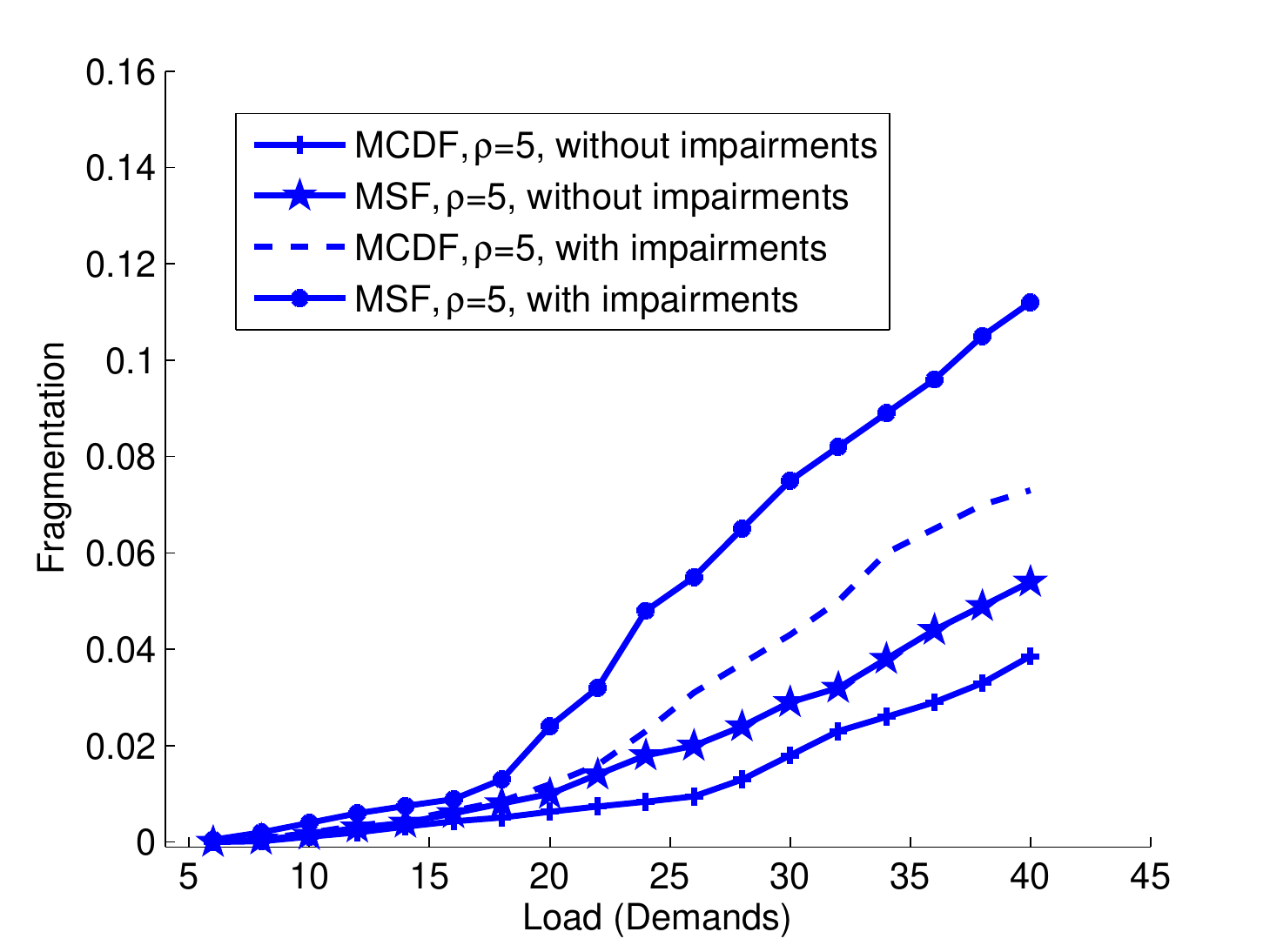}	\label{load_frag}} 
		\hspace{-.1cm}
		\subfigure[]{\includegraphics[trim={0cm 10cm 19cm 0cm}, height=.2\textwidth, width=.3\textwidth]{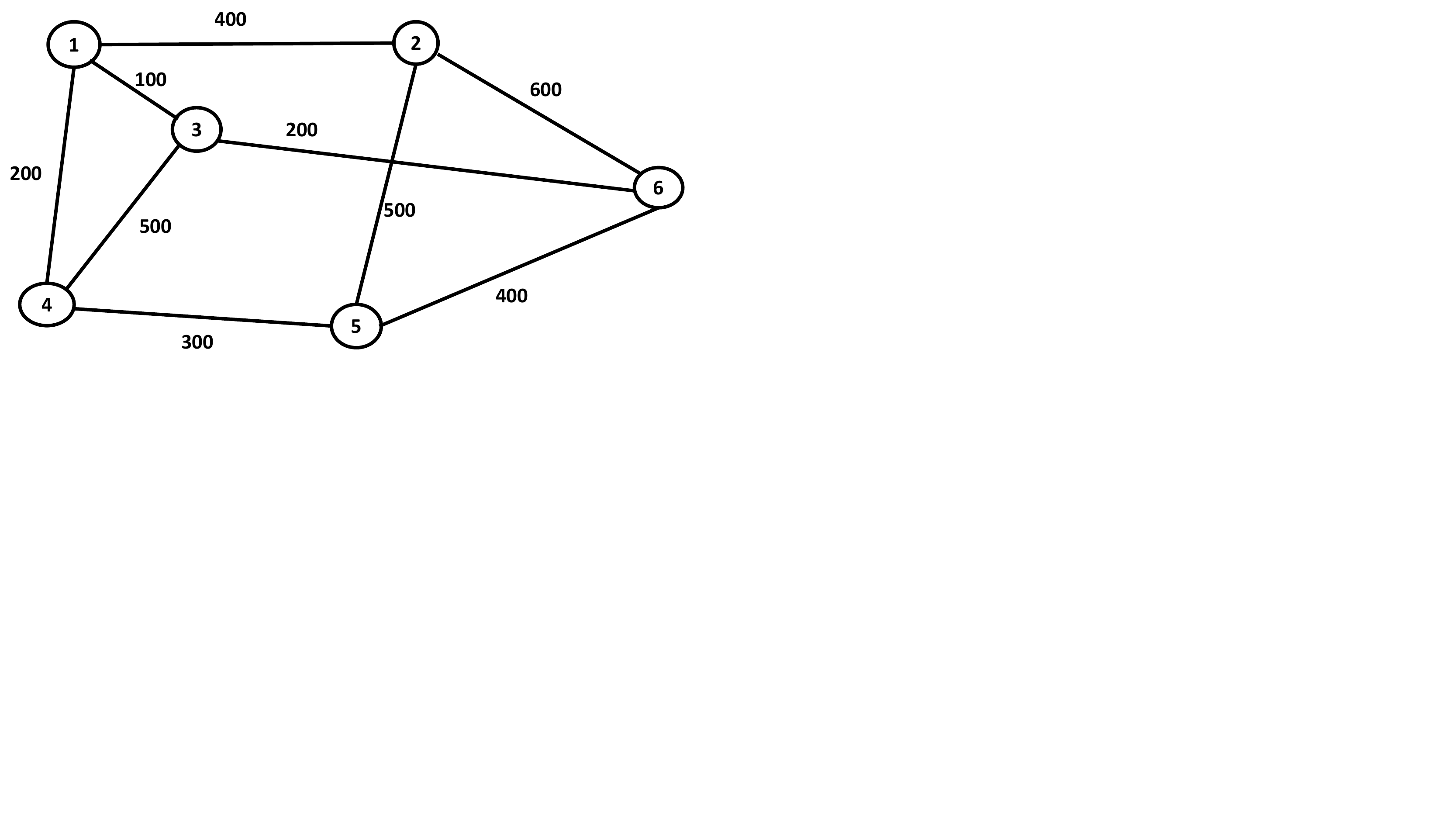}	\label{6-node}} 
		\caption{(a) Blocking probability for Heuristics with and without crosstalk for uniform bandwidth requests (b) Fragmentation for Heuristics  with and without crosstalk for uniform bandwidth requests (c) sample 6-node network.}
		\label{block}
	\end{center}
\end{figure*}
%\vspace{-.3cm}
\subsection{Comparison of DBP ILP with proposed ILP for Small Network}
The sample 6-node network considered for our study is shown in Fig. \ref{block}(c). The numbers on the link indicate link distance in Kms.
We have compared the DBP ILP with the proposed ILP in terms of fragmentation and delay-bandwidth product with and without impairments.
We define fragmentation  as follows \cite{wang2012spectrum},
\begin{align*}
F=1-({\text{largest   continuous   free   slots block}}/{\text{total  free slots}}) 
\end{align*}

Performance analysis in terms of fragmentation and delay-bandwidth product for sample 6-node network (Fig. \ref{block}(c))  is summarized in Table \ref{frag}.  $F_{avg}$ denotes average fragmentation considering all the links of a particular set of demands. We  observe that the proposed modified  ILP performs better than the DBP ILP in terms of fragmentation at the expense of marginal increase in delay-bandwidth product.  With impairments, the demands through  same XC are to be provided with separate slots to keep the accumulated crosstalk levels below the acceptable limit. This might create additional fragmentation. However, as we target to reduce fragmentation, a minimal increase in delay-bandwidth product is observable as the potential interference creator is shifted to its next shortest path available.

%\vspace{-.4cm} 
\begin{table}[H]
	\centering
	\caption{Fragmentation Analysis}
	\label{frag}
	\begin{tabular}{|c|c|c|}
		\hline
		System Model      & Without Impairments & With Impairments \\ \hline
		DBP ILP \cite{behera2016transmission}          & $F_{avg}$=18.04\%        & $F_{avg}$=20.26\%     \\ \hline
		Proposed ILP      & $F_{avg}$=3.86\%         & $F_{avg}$=4.48\%      \\ \hline
		Delay-bandwidth product   & 10.66\%              & 12.03\%           \\ 
		increase	for proposed ILP &  & 	\\	\hline
	\end{tabular}
\end{table}
We have compared the proposed RSA problem (ILP result) with and without impairments. On an average 33\% of the demands (for 6 node; 6, 8 and 10 demands) do not meet the SINR constraint when they are allocated without considering impairments and the deviation can be as worse as 2.8 dB. This shows essentiality of the impairment aware RSA as without it, many links might suffer from heavily degraded performance.
%\vspace{-.3cm}
\subsection{Comparison of ILP with Heuristics for Small Network} 
Performance analysis of the proposed ILP and the heuristics for 6-node  network with impairments is summarized in Table \ref{6-node table} in terms of simulation time (T) and optimality gap (G). The optimality gap  refers to the percentage of average deviation of a heuristic algorithm from the ILP in terms of the value of the overall objective function \eqref{mod_obj}.  We have considered demands (D) as 4 and 6 and $N=20$ slots per link. We observe that the ILP could not give solution for more than 6 demands when impairments are considered. 
%\vspace{-.3cm}
\begin{table}[h]
	\centering
	\caption{Performance analysis of small network}
	\label{6-node table}
	\begin{tabular}{|c|c|c|c|c|c|}
		\hline
		\begin{tabular}[c]{@{}c@{}}Load\\ (D)\end{tabular} & ILP         & MSF         & MCDF   & \multicolumn{2}{c|}{Optimality Gap G (\%)} \\ \hline
		& T {[}sec{]} & T {[}sec{]} & T {[}sec{]} & ILP-MSF           & ILP-MCDF          \\ \hline
		4                                                  & 600         & 0.11        & 0.13        & 8.1              & 2.77                   \\ \hline
		6                                                  & 3600        & 0.15        & 0.18        & 13.15             & 3.45                   \\ \hline
	\end{tabular}
\end{table}
%\vspace{-.35cm}
\subsection{Comparison of Heuristics for Realistic Networks}
We have evaluated the proposed algorithms for 14-node NSFNET \cite{wang2012spectrum} since our ILP is computationally more intensive for generating results of larger networks.Performance analysis of heuristics for NSFNET along with impairments are summarized in Table \ref{14-node table} in terms of simulation time (T) and objective function value (Obj). We have considered $N=40$ slots per link. 
%\vspace{-.4cm}
\begin{table}[]
	\centering
	\caption{Performance analysis of realistic network}
	\label{14-node table}
	\begin{tabular}{|c|c|c|c|c|}
		\hline
		\begin{tabular}[c]{@{}c@{}}Load \\ (D)\end{tabular} & \multicolumn{2}{c|}{MSF} & \multicolumn{2}{c|}{MCDF} \\  \hline
		& T {[}sec{]}   & Obj      & T {[}sec{]}      & Obj         \\  \hline
		6                                                   & 0.21         & 65.37  & 0.24             & 57.61      \\  \hline
		10                                                  & 0.24          & 147.18   & 0.27             & 125.4      \\  \hline
		15                                                  & 0.29          & 262.4   & 0.33             & 211.66     \\ \hline
		20                                                  & 0.34          & 381.26    & 0.39             & 334.18      \\  \hline
	\end{tabular}
\end{table}

%\begin{figure*}[t!]
%	%	\centering
%	\begin{center}
%		\vspace{-0.5cm}
%		\hspace{-1.5cm}	
%		\subfigure[]{\includegraphics[height=.3\textwidth, width=.33\textwidth]{load_bbp}\label{load_bbp}} 
%		%		\hspace{-2cm}
%		%		\captionsetup{justification=centering,margin=2cm}
%		\subfigure[]{\includegraphics[height=.3\textwidth, width=.33\textwidth]{load_frag}	\label{load_frag}} 
%		\hspace{-.1cm}
%		\subfigure[]{\includegraphics[trim={0cm 10cm 19cm 0cm}, height=.2\textwidth]{6-node}	\label{6-node}} 
%		\caption{(a) Blocking probability for Heuristics with and without crosstalk for uniform bandwidth requests (b) Fragmentation for Heuristics  with and without crosstalk for uniform bandwidth requests (c) sample 6-node network}
%	\end{center}
%\end{figure*}
From Table \ref{14-node table}, we can observe that MCDF performs better than MSF at the expense of marginal increase in simulation time due to its different ordering than MSF. 
We have compared our multi objective formulation with existing BLSA scheme as mentioned in the introduction. BLSA uses MSF ordering  without impairments, so we have used MSF ordering without impairments  for comparison purpose.  We found that, there is 19.62\% increase in delay-bandwidth product in BLSA scheme at the cost of 2.27\% increase in fragmentation for our formulation. The maximum subcarrier index (MS) for BLSA is 14 on any link as opposed to 17 for our case. The allocation in BLSA is insensitive to delay as it does load balancing from the beginning of spectrum allocation which is shown in Table \ref{comparison}.  
% \vspace{-.4cm}
\begin{table}[]
	\centering
	\caption{Comparison with \cite{wang2011study}}
	\label{comparison}
	\begin{tabular}{|c|c|c|}
		\hline
		& BLSA    & Proposed Heuristic \\ \hline
		MS                        & 14      & 17                 \\ \hline
		Ordering technique        & MSF     & MSF                \\ \hline
		Increase in delay-bandwidth         & 19.62\% &  0 \%                  \\ \hline
		Increase in fragmentation &  0 \%       & 2.27\%             \\ \hline
	\end{tabular}
\end{table}

Fig. \ref{block}(a) shows  blocking probability for proposed ordering techniques with and without impairments for different loads. We have considered total $N=40$ slots and $K=3$ paths per connection with uniform bandwidth requests of $\rho=5$ to encourage blocking.    We can observe that the blocking probability increases with impairments and MCDF performs better than MSF for both with and without impairments.  Similarly, MCDF outperforms MSF in terms of fragmentation as demonstrated  in Fig. \ref{block}(b). We can observe that without impairments, the performance of MCDF and MSF are comparable as the load increases. However, with impairments MCDF performs better than MSF which signifies the importance of considering physical layer impairments in the network layer spectrum allocation.

\section{Conclusion}
In this paper, we have addressed the RSA problem that jointly optimizes the delay-bandwidth product, fragmentation and congestion along with transmission impairments such as in-band crosstalk generated due to cross-connect, shot noise, ASE and crosstalk beat noise due to coherent reception, nonlinear interference and filter narrowing. We have formulated an ILP  and proposed two different demand ordering techniques for the heuristic.  We first validate the performance of our proposed algorithms with the proposed ILP for a small network.  We have calculated fragmentation and observed that the proposed ILP achieves significantly lesser fragmentation compared to that of DBP optimization framework  at the expense of marginal increase in delay-bandwidth product. Furthermore, we have compared our algorithm with the existing BLSA scheme and observe that our formulation achieves better performance in terms of delay-bandwidth product at the expense of marginal increase in fragmentation. We have further compared the two different ordering algorithms  for a realistic network in terms of fragmentation and blocking probability.   Through simulations, we demonstrate  the effect of impairments in an RSA and show that neglecting impairments can severely degrade the quality (SINR) of most of the optical links.

\appendices
\numberwithin{equation}{section}
%\vspace{-0.2cm}
\section{Probability of symbol error  for 4-QAM Heterodyne Receiver}
\begin{figure*}
	\centering
	\includegraphics[trim={0cm 5cm 0cm 0cm}, width=1\textwidth, height=.35\textwidth]{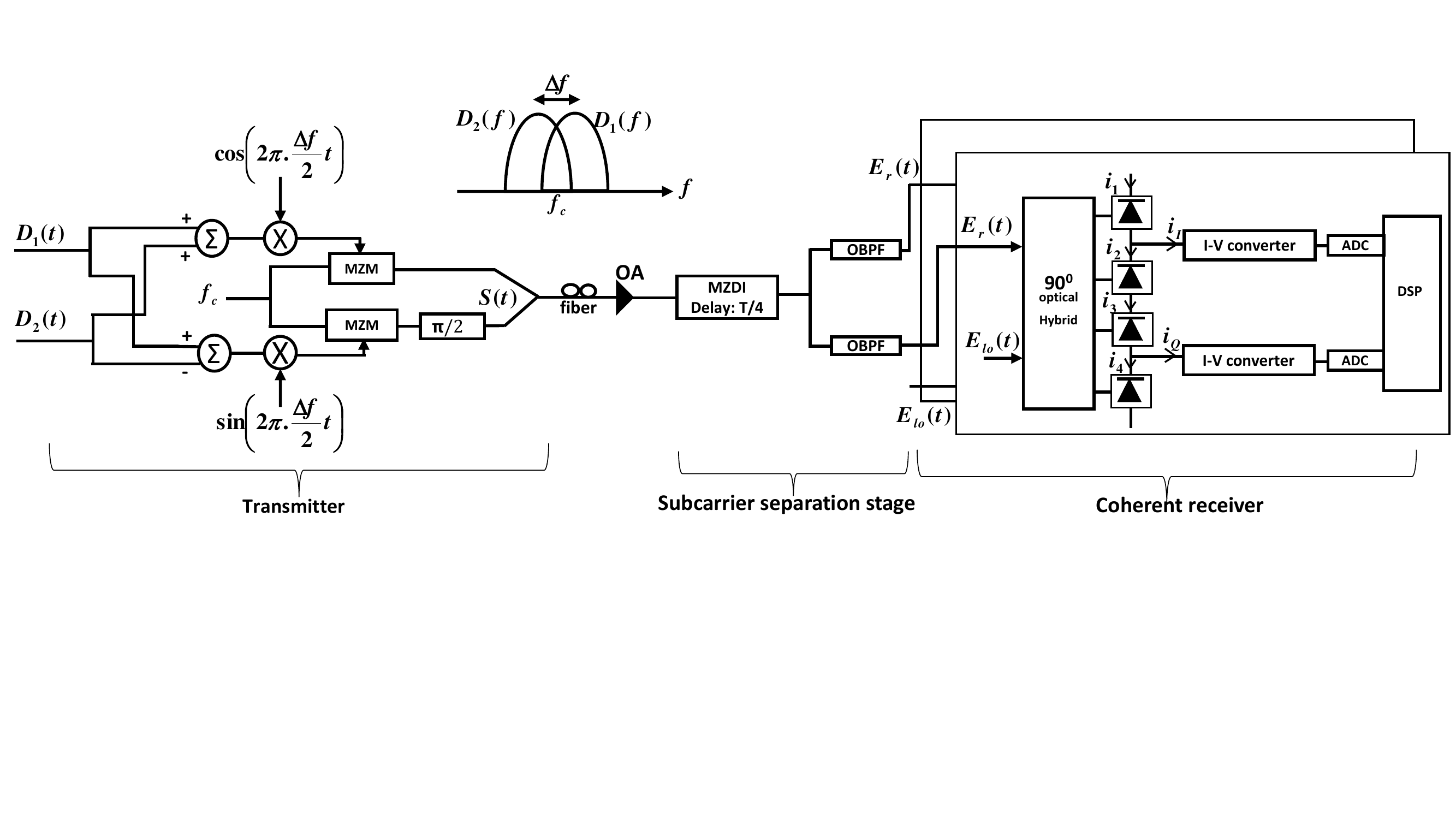}	
	\caption{Electro-optical CO-OFDM employing 2 subcarriers; MZM: Mach-Zehnder modulator, MZDI: Mach-Zehnder delay interferometers, OA: Optical amplifier, OBPF: Optical band pass filter}
	\label{OFDM}
\end{figure*}
 Electro-optical CO-OFDM with two subcarriers is shown in Fig. \ref{OFDM}, where $D_1(t)$ and $D_2(t)$ are modulated data patterns (for our case 4-QAM samples), $\Delta f$ is the frequency spacing of the two subcarriers \cite{kobayashi2009over,sanjoh2002optical,sano2009no}. This can be easily generalized for any number of subcarriers. Balanced heterodyne and homodyne  detection are the two possible detection processes for coherent  systems. Since  EON employs coherent optical OFDM, it is necessary to select either of these detection processes. Though the homodyne detection has improved sensitivity over heterodyne detection, the implementation of the homodyne detection   is more complex.

\textit{\textbf{Assumptions:}} In our model, we  assume a balanced heterodyne detection with each subcarriers employing a 4-QAM modulation. Further, we assume that all the losses are exactly compensated by  Erbium Doped Fiber Amplifier (EDFA). In addition, we consider the primary signal and the interfering signal are having same attributes in-terms of identical bit rate, modulation format (as a consequence identical PSD is assumed).  Therefore, we assume that at each XC, every outgoing lightpath gets exposed to same amount of interference power from other interfering lightpaths (both in-band and out-of-band). For simplicity, we have assumed the crosstalk factor to be constant when the signal traverses through multiple XCs and have considered a high value (-18.5 dB) for worst case design. 
 We assume that laser has no drift, Mach-Zehnder delay interferometers (MZDI) are perfectly synchronized. Crosstalk and losses associated with all components are neglected and all subcarriers are exposed to same level of interference. Hence, we derive the bit error rate (BER) performance for one subcarrier like a single-carrier system. 

%\begin{figure*}
%	\centering
%\includegraphics[trim={0cm 5cm 0cm 1cm}, width=1\textwidth, height=.35\textwidth]{OFDM.pdf}	
%\caption{Electro-optical CO-OFDM employing 2 subcarriers; MZM: Mach-Zehnder modulator, MZDI: Mach-Zehnder delay interferometers, OA: Optical amplifier, OBPF: Optical band pass filter}
%\label{OFDM}
%\end{figure*}

 In addition to providing gain to the attenuated signal, optical amplifier (OA) also introduces ASE noise. The detection performed at the receiver is Balanced Coherent Optical Detection (BCOD), which consists of an optical hybrid and photo diode arrangement for balanced detection. The transmitted optical signal after amplification and polarization control is allowed to mix with LO optical signal in the $90^{0}$ optical hybrid. The I channel and Q channel photocurrents from the BCOD are current to voltage (I-V) converted  and then processed in Digital Signal Processor (DSP) (Fig. \ref{OFDM}).     

\begin{figure}[H]
	\centering
	\includegraphics[trim={0cm 10cm 5cm 0cm}, width=.5\textwidth]{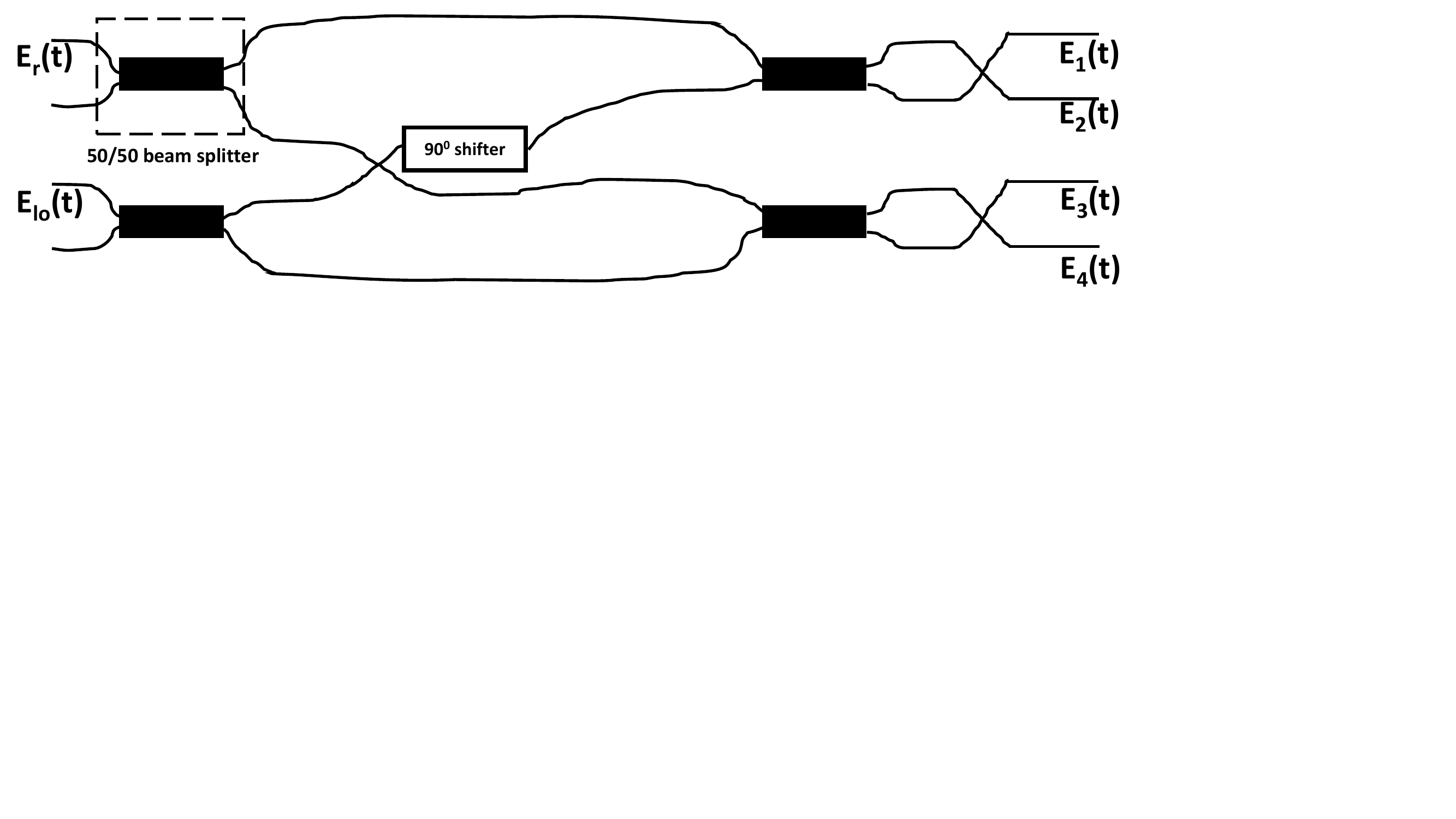}
			\caption{${90}^0$ Optical hybrid}
			\label{opt}
\end{figure}

Detailed setup of the ${90}^0$ optical hybrid is schematically shown in Fig. \ref{opt}. 
Each individual splitter is called 50/50 beam splitter or ${180}^0$ hybrid. 

The received electric field relationship in four arms of the ${90}^0$ optical hybrid are,
\vspace{-.2cm}

\footnotesize
\begin{align}
\label{eqn:E1}
 &\vert \bm{E_{1}(t)}\vert^2 =  \vert\frac{1}{2}j \left(\bm{E_{r}(t)}+\bm{E_{lo}(t)}\right) |^2={|\frac{1}{2}\left(\bm{E_r(t)}+\bm{E_{lo}(t)}\right)\vert}^2,\nonumber \\
%\label{eqn:E2}
&\vert\bm{ E_{2}(t)}\vert^2 =  |\frac{1}{2}\left(\bm{E_{r}(t)}-\bm{E_{lo}(t)}\right) |^2={|\frac{1}{2}\left(\bm{E_r(t)}-\bm{E_{lo}(t)}\right)\vert}^2, \nonumber \\
%\label{eqn:E3}
&\vert \bm{E_{3}(t)}\vert^2 =  |\frac{1}{2}j\left(\bm{E_{r}(t)}+j\bm{E_{lo}(t)}\right) |^2={|\frac{1}{2}\left(\bm{E_r(t)}+j\bm{E_{lo}(t)}\right)\vert}^2,\nonumber \\
%\label{eqn:E4}
&\vert \bm{E_{4}(t)}\vert^2 =  |\frac{-1}{2}\left(\bm{E_{r}(t)}-j\bm{E_{lo}(t)}\right) |^2={|\frac{1}{2}\left(\bm{E_r(t)}-j\bm{E_{lo}(t)}\right)\vert}^2. 	
\end{align}
\normalsize

\noindent These complex electric fields are detected by PIN photodiodes (PD). The optical power received ($P_k (t)$), and the corresponding photocurrents ($i_k (t)$) generated in each PD are given as follows,
\vspace{-.4cm}

\footnotesize
\begin{align}
\label{eqn:Pkinitial}
&P_k (t)=|\bm{E_k (t)}|^2, \\
&i_k (t)=R_{a}P_k(t)=R_{a}|\bm{E_k (t)}|^2,
\end{align}\normalsize

\noindent where \textit{k}= 1, 2, 3, 4 and we assume that responsivity ($R_a$) is same for all the four photodiodes.
The in-phase and quadrature-phase currents generated in the balanced detector are,
\vspace{-.4cm}

\footnotesize
\begin{align*}
i_I(t)= i_1(t) - i_2(t),\\
i_Q(t)= i_3(t) - i_4(t).
\end{align*}\normalsize

\noindent An LO laser is required to beat with the received signal to recover the signal amplitude and phase information.
Let the LO electric field and the received signal after polarization control be,

\vspace{-.4cm}

\footnotesize
\begin{align}
&E_{lo}(t)=\sqrt{P_{lo}} e^{j\phi_{lo}(t)}e^{j\omega_{lo}t}, \nonumber \\
& E_r(t)=\left[\sqrt{P_{r}(t)}e^{j( \theta_{s}(t) + \phi_{s}(t))} + \overline{n}_x(t)\right] e^{j\omega_c t}
\label{eqn:Er}
\end{align}\normalsize

\noindent where $P_r(t)$ and $\theta_s(t)$ are the power and the phase of the transmitted symbol, $P_{lo}(t)$ is the LO power,  $\phi_s(t)$ and $\phi_{lo}(t)$  are the phase noise associated with the source laser and LO; $\omega_c$ and $\omega_{lo}$  are the carrier frequency of the signal and LO respectively, $\overline{n}_x(t)$ is the complex envelope (low pass representation) of ASE noise. We assume $\phi_s(t)$  is slow varying (i.e., $\phi_s(t)$ is at least non varying over more than one symbol duration) and $P_r(t)$ is constant over a symbol duration for a QAM signal. 

If the LO is phase synchronized with the incoming data signal (i.e, $\phi_{lo}(t)$ = $\phi_{s}(t)$), 
\vspace{-.4cm}	

\footnotesize
\begin{align}
\label{currI}
&i_I(t)=I_I(t)+i_{I{n}}(t)=R_{a}\sqrt{P_r P_{lo}}cos\left(\omega_{if}t+\theta_s(t)\right)+i_{I{n}}(t), \nonumber \\	
&i_Q(t)=I_Q(t)+i_{Q{n}}(t)=R_{a}\sqrt{P_r P_{lo}}sin\left(\omega_{if}t+\theta_s(t)\right)+i_{Q{n}}(t)
\end{align}\normalsize

\noindent where, $\omega_{if}=\omega_{c}-\omega_{lo}$ is the intermediate frequency (IF). $i_{I{n}}(t)$ and $i_{Q{n}}(t)$ are I channel and Q channel noise currents respectively.
	
	The autocovariance of photocurrent I(t) is defined as \cite{ramaswami2009optical},
	\vspace{-.4cm}	
	
	\footnotesize
	\begin{align}
	\label{eqn:autocov}
	L_I(\tau)&= E\left[I(t)I(t+\tau)\right]-E\left[I(t)\right]E\left[I(t+\tau)\right] \nonumber \\&=E\left[ P(t)\right]eR_{a}\delta(\tau) + {R_{a}}^2L_P(\tau),
	\end{align}\normalsize
	
	\noindent where $L_P(\tau)$ is the auto co-variance of the received optical power.
	The expression represented by \eqref{eqn:autocov} is used in deriving the auto co-variance of I and Q channel currents in this paper.
	\vspace{-.6cm}
	
	 \subsection{Noise Statistics of Optical Heterodyne Detection}
	 In heterodyne detection, the LO frequency ($f_{lo}$) is not same as the incoming data signal frequency ($f_{c}$). We assume that the detection is performed at perfect synchronization with the incoming signal. We derive for noise only to avoid complexity of representation and later on we take interference into account.
	 Let the complex envelop of the ASE noise be, 
	 
%		 \footnotesize
\begin{equation}
% \begin{align}
	 \label{eqn:nx}
	 \overline{n}_x(t)=x(t)+j \, y(t),
%	 \end{align}
	 \end{equation}
%	 \normalsize
	 
	 \noindent By combining \eqref{eqn:E1} and putting the values of $\vert \bm{E_{k}(t)}\vert^2$ with the values of  $E_r(t)$ and $E_{lo}(t)$ from \eqref{eqn:Er} in \eqref{eqn:Pkinitial}, we get the expressions of optical power present in four output branches of the $90^{o}$ optical hybrid with assumed constant LO power as,
	 \vspace{-.4cm}
	 
	 \footnotesize
	 \begin{align}
	 \label{eqn:Pkhetero}
	 &P_k(t)=\frac{1}{4}\Big[P_r  +|\overline{n}_x(t)|^2 + P_{lo}+ 2\sqrt{P_r}n_b(t)+2{(-1)}^{(k-1)}.\nonumber \\&\left(\sqrt{P_{lo}}n_{if}(t)+\sqrt{P_r P_{lo}}cos\left( \omega_{if}t+\theta_s(t)+\left \lfloor{\frac{k-1}{2}}\right \rfloor.\frac{\Pi}{2}\right)\right)\Big], 
	 \end{align}\normalsize
	 
	 \noindent where, k= 1, 2 , 3 and 4 corresponds to four branches in the $90^{0}$ optical hybrid and 
	 $\left \lfloor{u}\right \rfloor $ is the floor function of u; $n_{if}(t)$, $n_b(t)$ are the narrow band-pass ASE noise centered at intermediate frequency and the base band ASE noise respectively. The expression for  $n_{if}(t)$ and $n_b(t)$ are as follows:
	 
	 \noindent 
	 \vspace{-.6cm}
	 
	 \footnotesize
	 \begin{align*}
	 &n_{if}(t)=x(t)cos\left(\omega_{if}t-\phi_{lo}(t)\right)-y(t)sin\left(\omega_{if}t-\phi_{lo}(t)\right), \\&n_{b}(t)=x(t)cos\left(\theta_{s}(t)+\phi_{lo}(t)\right)-y(t)sin\left(\theta_{s}(t)+\phi_{lo}(t)\right).      
	 \end{align*}\normalsize

	 \noindent The current detected $i_k(t)$ in each arm can be written as the sum of the ideal current $I_k(t)$ (without noise) and the noise current $i_{kn}(t)$,
	 %Where $I_k(t)$ is current without noise and  $i_{kn}(t)$ is noise current.
	 \vspace{-.4cm}
	 
	 \footnotesize
	 \begin{align}
	 \label{eqn:ik}
	 i_k(t)&=I_k(t)+i_{kn}(t), \qquad  k=\text{1, 2, 3, 4}\\
	 \label{eqn:Ik}
	 I_k(t)&=\frac{R_{a}}{4}\Big[  P_r + P_{lo}+\nonumber \\& 2{(-1)}^{(k-1)}\sqrt{P_r P_{lo}}cos\left( \omega_{if}t+\theta_s(t)+\left \lfloor{\frac{k-1}{2}}\right \rfloor\frac{\Pi}{2}\right)\Big].    
	 \end{align}\normalsize

	 The auto co-variance of  $P_k (t)$ \eqref{eqn:Pkhetero} over a symbol interval is given by,	 
	 \footnotesize
	 \begin{align*} 
	 &L_{P_k}(\tau)=\frac{1}{16}\Big[4P_r R_{n_b}(\tau)+ 4P_{lo}R_{n_{if}}(\tau)+8{R^2_{n_{b}}(\tau)}\Big],\\
	 &L_{P_k}(\tau)=L_{P_j}(\tau) \qquad \forall \; j\; and \; k. \nonumber
	 \end{align*}\normalsize
	 
\noindent	where for a stationary Gaussian noise process, the auto-correlation of $n_{b}(t)$ and $n_{if}(t)$ are given as follows,
	  \vspace{-.4cm}
	  
	  \footnotesize
	  \begin{align*}
	  &R_{n_{b}}(\tau)=R_{xx}(\tau),	\\
	  &R_{n_{if}}(\tau) =R_{xx}(\tau)cos(\omega_{if}\tau)+R_{yx}(\tau)sin(\omega_{if}\tau),		               
	  \end{align*}\normalsize
	  
	  \noindent where, $R_{xx}(\tau)$ is the  autocorrelation of the in phase component [$x(t)$] of the ASE noise, and $R_{yx}(\tau)$ is the cross correlation of the quadrature component [$y(t)$] and the in phase component  [$x(t)$] of the ASE noise.

%	 \noindent Evaluating \eqref{eqn:Pkhetero} on \eqref{eqn:autocov} with the substitution $P(t)=P_k(t)$ and $L_{P}(\tau)=L_{P_k}(\tau)$, we obtain the corresponding auto co-variance of noise currents as, 
	 
	\noindent We compute auto co-variance of \eqref{eqn:ik} using \eqref{eqn:autocov} where $P(t)$ is substituted with $P_k(t)$ from \eqref{eqn:Pkhetero}, as follows:
	 \vspace{-.4cm}
	 
	 \footnotesize
	 \begin{align*}
	 L_{i_k}(\tau)=\frac{eR_{a}}{4}\big[ P_r +P_{lo}+R_{n_{b}}(0)\big]\delta(\tau) +\frac{R_{a}^2}{16}\big[4P_r R_{n_b}(\tau)\nonumber \\ + 4P_{lo}R_{n_{if}}(\tau)+8{R^2_{n_{b}}(\tau)}\big],
	 \end{align*}\normalsize
	 
	 \noindent where, $k=1,2,3,4$ corresponding to each arms and $R_{n_{b}}(0)$ is the ASE noise power.
	 \vspace{-.4cm}
	 
	 \footnotesize
	 \begin{align}
	 \qquad R_{n_b}(0)= n_{sp}(G-1)hf_cB_o.
	 \end{align}\normalsize

	 \noindent We assume noise in the two arms of balanced detector are uncorrelated, then the auto co-variance of I channel noise current,
	 \vspace{-.4cm}
	 
	 \footnotesize
	 \begin{align}
	 \label{fo}
  	      \hspace{-.3cm}L_{i_I}(\tau)=\frac{eR_{a}}{2}\left[P_r +P_{lo}+R_{n_{b}}(0)\right]\delta(\tau) +\frac{R_{a}^2}{2}\big[P_r R_{n_b}(\tau)  \nonumber \\ + P_{lo}R_{n_{if}}(\tau)+2{R^2_{n_{b}}(\tau)} \big].
	 \end{align}\normalsize
	 
	 \noindent Similarly the auto co-variance of Q channel noise current,
	 \vspace{-.4cm}
	 
	 \footnotesize
	 \begin{align*}
	 L_{i_Q}(\tau)=L_{i_I}(\tau).
	 \end{align*}\normalsize
	 
	 Therefore, the power spectral density of the I channel noise current is (taking the Fourier transform of \eqref{fo}),
	 \vspace{-.4cm}
	 
	 \footnotesize
	 \begin{align}
	 \label{psdhetero}
	 S_{I_I}(f)=\frac{eR_{a}}{2}\left[P_r+P_{lo}+2R_{n_{b}}(0)\right]+\frac{R_{a}^2}{2}P_rS_{N_b}(f) \nonumber \\+ \frac{R_{a}^2}{2}P_{lo}S_{N_{if}}(f)+4R_{a}^2{S_{N_{b}}(f)}\ast{S_{N_{b}}(f)},
	 \end{align}\normalsize
	 
	 \noindent where $S_{N_{if}}(f)$ is the PSD of narrow band pass ASE noise centered at IF ($n_{if}(t)$) and $S_{N_{b}}(f)$ is the PSD of low pass ASE noise ($n_{b}(t)$). It is clear that the power spectral density of the Q channel noise current is same as that of the I channel, for the same received optical signal power $P_r$, because they have the same auto covariance.
	 \vspace{-.4cm}
	 
	 \footnotesize
	 \begin{align}
	 S_{I_Q}(f)=S_{I_I}(f).
	 \end{align}\normalsize
	  
	  \noindent From the above derived PSD, the term $ \frac{eR_{a}}{2}(P_r+P_{lo}+2R_{n_{b}}(0))$ represents shot noise due to photo detection process and other terms represent ASE beating noise products. The dominant noise term $\frac{R_{a}^2}{2}P_{lo}S_{N_{if}}(f)$ is the local oscillator-ASE noise beat product. The other term like $\frac{R_{a}^2}{2}P_r S_{N_b}(f)$ represents the signal-ASE noise beat product and the last term $4R_{a}^2{S_{N_{b}}(f)}\ast{S_{N_{b}}(f)}$ corresponds to the ASE-ASE beat noise.
	 
	 To account for crosstalk, the received electric field of the lightwave after the polarization control and the dispersion control  can be written as,
	 \vspace{-.4cm}	 
	 
	  \footnotesize
	  \begin{align}
	  \label{eqn:ErOFDM}
	  E_r(t)=&\left[\sqrt{P_{r}(t)}e^{j( \theta_{s}(t) + \phi_{s}(t))} + \overline{n}_x(t)\right] e^{j\omega_c t} \; \nonumber\\&+   \sum_{l} \left[ \sqrt{P_{Xtalk_{l}}+Pnli_{l}}e^{j( \theta^\prime_{s_{l}}(t) + \phi^\prime_{s_{l}}(t))} \right] e^{j\omega_c t} 
	  \end{align}
	  \normalsize
	 
	\noindent where $P_{Xtalk_{l}}$ and $\theta^\prime_{s_{l}(t)}$ are the power and the signal phase of the interfering in-band signal, $\phi^\prime_{s_{l}(t)}$ is the phase noise associated with the '$l^{th}$ 'interfering source laser. $Pnli_{l}$ is the NLI of '$l^{th}$ 'interfering source laser having center frequency $\omega_c$.
	By following similar procedure we derive PSD of $i_{I}(t)$ and $i_{Q}(t)$  as given in \eqref{psdhetero}. 
	 
	 The balanced coherent detection yields the photo currents in both the I arm and Q arm.
	 The generated photo current in the I channel (same as Q channel) can be written as,
%	 \vspace{-.45cm}	 
	 
	  \footnotesize
	  \begin{align}
	  \label{iIOFDM}
	  i_{I}(t)=&R_{a}\sqrt{P_{r}P_{lo}} cos\left(\omega_{if}t+\theta_s t\right)+i_{I\; {s-sp}}(t)+ i_{I\; {lo-sp}}(t) \;\nonumber \\ & +  i_{I \;{sp-sp}}(t) +  \sum_{l} i_{I\; {s-x_{l}}}(t) + \sum_{l} i_{I\; { lo-x_{l}}}(t) \;\nonumber \\ & + \sum_{l} i_{I\; {s-x_{nli}}}(t) + \sum_{l} i_{I\; { lo-x_{nli}}}(t)+\sum_{l} i_{I\; { x_{nli}-x_{nli}}}(t) \;\nonumber \\ & + \sum_{k,l} i_{I\; {x_{k}-x_{l}}}(t)   + i_{I\; shot}(t) 
	  \end{align}
	  \normalsize
		 The first part is the primary signal part, and the remaining terms are the noise, crosstalk and NLI terms. The noise terms are explained below as follows;
	 $i_{I\; s-sp}(t)$: primary signal-ASE beat noise,
	 $i_{I\; {lo-sp}}(t)$: LO signal-ASE beat noise,
	 $i_{I\; {sp-sp}}(t)$: ASE-ASE beat noise,
	 $\sum_{l} i_{I\; {s-x_{l}}}(t)$: sum of all the primary signal-cross-talk beat noises,
	 $\sum_{l} i_{I\; {lo-x_{l}}}(t)$: sum of all the LO-crosstalk beat noises,
	 $\sum_{l} i_{I\; {s-x_{nli}}}(t)$: sum of all the primary signal-NLI beat noises,
	 	$ \sum_{l} i_{I\; { lo-x_{nli}}}(t)$: sum of all the LO-NLI beat noises,
	 	$ \sum_{l} i_{I\; { x_{nli}-x_{nli}}}(t)$: sum of all the NLI-NLI beat noises, 
	 $\sum_{k,l} i_{I\; {x_{k}-x_{l}}}(t)$: sum of all the crosstalk-crosstalk beat noises,
	 $i_{I \;shot}(t)$: shot noise current.
		 
 The combined noise current can be modeled as a zero-mean Gaussian random process with a variance given by,

	 \footnotesize
	 \begin{align}
	 \label{sigmaOFDM}
	 \sigma_{i}^2= \sigma_{i \;s-sp}^2 + \sigma_{i \; lo-sp}^2 +\sigma_{i sp-sp}^2 +  \sigma_{i \; s-x}^2  +\sigma_{i \; lo-x}^2 \nonumber \\+ \sigma_{i \; s-nli}^2 +\sigma_{i \; lo-nli}^2  +\sigma_{i \; nli-nli}^2 +\sigma_{i \; x-x}^2 + \sigma_{i \; shot}^2
	 \end{align}
	 \normalsize
	 
	  \noindent Since we assumed heterodyne detection, the noise variance for I-channel and and Q-channel are same. The variances $\sigma_{i\;  sp-sp}^2$, $\sigma_{i \; x-x}^2$, $\sigma_{i \; s-nli}^2$, $\sigma_{i \; nli-nli}^2$ and $\sigma_{i \; s-x}^2$ are negligible comparing to other noise variances.
	 The corresponding noise variances are derived as,

	\footnotesize
	\begin{align}
	& \hspace{-1cm} \sigma_{i \;s-sp}^2 =\frac{R_{a}^2}{2} P_r n_{sp}hf_c (G-1) B_{e} = \frac{R_{a}^2}{2} P_r S_{ASE} B_{e} \nonumber \\
	& \hspace{-1cm}\sigma_{i \; lo-sp}^2=\frac{R_{a}^2}{2} P_{lo} n_{sp}hf_c (G-1) B_{e} = \frac{R_{a}^2}{2} P_{lo} S_{ASE} B_{e} \nonumber \\
	& \hspace{-1cm}\sigma_{i \; lo-x}^2=\frac{R_{a}^2}{2}\sqrt{P_{lo}P_{xtalk_l}
	} \nonumber \\
	& \hspace{-1cm}\sigma_{i \; lo-nli}^2=\frac{R_{a}^2}{2}\sqrt{P_{lo}Pnli{_l}
		} \nonumber \\
		& \hspace{-1cm} \sigma_{i \; shot}^2=\frac{eR_{a}}{2}\left[ P_r +P_{lo}+ \sum_{l}P_{xtalk_{l}}+2R_{n_{b(0)}}\right]    
		\label{sig}
		\end{align}
		\normalsize	
	  
\noindent where $P_{xtalk_l}$ is calculated using \eqref{xt} and $Pnli{_l}$ is calculated according to \eqref{NL}, \eqref{NLI1}, and \eqref{NLI2}.
	
	 In order to find I-channel and Q-channel QAM signal points, first to let the photo current to pass through IF filter having bandwidth $B_{IF}$($<<B_o$, Optical bandwidth), next we multiply (\ref{iIOFDM}) with $cos\left(\omega_{if}t\right)$ and then average over one symbol period. We assume  $B_{IF}=2B_e$, where $B_e$ is the electronic bandwidth. The IF filter rejects the out-of-band noise signals. The resulted signal points  are given as,
	 \vspace{-.4cm}	 
	 
	 \footnotesize
	 \begin{align*}
	 &i'_{I(t)}=E[i_I(t) \times cos\left(\omega_{if}t\right)]=I'_I(t) +i'_{In}(t), \\
	 &i'_{Q(t)}= E[I_Q(t) \times cos\left(\omega_{if}t\right)]=I'_Q(t) +i'_{Qn}(t),
	 \end{align*}
	 \normalsize
	 
	 \noindent where $i'_{In}$ and $i'_{Qn}$ are the received I-channel and Q-channel constellation points at the decision making point, with mean $I'_I(t)$ and $I'_Q(t)$ respectively, and variances $\sigma^2_I$ and $\sigma^2_Q$ respectively.
	 \vspace{-.4cm}	 
	 
	 	 \footnotesize
	 \begin{align}
	 \label{Icurrenthetero}
	 &I'_I(t)= R_{a}\sqrt{P_r P_{lo}} cos\left(\theta_s(t)\right)/2,\\
	 \label{Qcurrenthetero}
	 &I'_Q(t)=R_{a}\sqrt{P_r P_{lo}} sin\left(\theta_s(t)\right)/2
	 		 \end{align}
	 \normalsize
	 
 The variance of I-channel (same as Q channel) noise current is given in (\ref{sigmaOFDM}) and (\ref{sig}). 
	Now we intend to evaluate BER performance of coherently received  optical 4-QAM signal.  We assume the minimum distance between two constellation points to be $2d$. Four symbols are provisioned in 4-QAM with assumed the powers for each symbol being $P_{r1}$, $P_{r2}$, $P_{r3}$ and $P_{r4}$. However, for 4-QAM  $P_{r}=P_{r1}=P_{r2}=P_{r3}=P_{r4}$. Where \text{from \eqref{Icurrenthetero}-\eqref{Qcurrenthetero}}.,
	\vspace{-.4cm}	 
	
		 \footnotesize
	 \begin{align}
	 d= \frac{R_{a}}{2} \sqrt{P_{r}P_{lo}/2} 
	 \label{dval}
	 \end{align}
	 	 \normalsize
	 	  The coherently detected signal's power $(P_{ch})$ is given as,
	 	  \begin{align}
	 	  %P_{ch}=\frac{R_{a}^2}{2}P_{lo}P_{r}
	 	  P_{ch}=({R_{a}^2}/{2})P_{lo}P_{r}
	 	  \label{pch}
	 	  \end{align}
	 	  For high local oscillator (LO) power, shot noise and all other beat noise terms can be neglected except 
	 	  \vspace{-.2cm}
	 	  \begin{align}
	 	  & \hspace{-1cm}\sigma_{i \; lo-sp}^2=({R_{a}^2}/{2}) P_{lo} M_k n_{sp}hf_c (G-1) B_{e} \label{sinr1} \\
	 	  & \hspace{-1cm}\sigma_{i \; lo-x}^2=\frac{R_{a}^2}{2}\sqrt{P_{lo}P_{xtalk_{l}}} 
	 	  \label{sinr2} \\
	 	  & \hspace{-1cm}\sigma_{i \; lo-nli}^2=\frac{R_{a}^2}{2}\sqrt{P_{lo}Pnli{_l}
	 	  	}
	 	  \end{align}
	 	  
	 	  \noindent where  $M_k$ is the number of EDFA's in the $k$-th lightpath.

	 If all the symbols are equiprobable, then the average signal power for 4-QAM constellation at the decision point,
	 \vspace{-.4cm}
	 
	 \footnotesize
	 \begin{align*}
	 P_{avg_{4-QAM}}=d\sqrt{2}.
	 \end{align*}
	 \normalsize
	 
	 \noindent The probability of symbol error for 4-QAM,
	 \vspace{-.3cm}
	 	 
	 \footnotesize
	 \begin{align}
	 P_e= &1-\left[1-Q\left( \frac{d}{\sigma_{I}}\right) \right] \times \left[ 1- Q\left( \frac{d}{\sigma_{Q}}\right) \right] \nonumber \\=&2Q\left( \sqrt{\frac{d^2}{\sigma^2_{I}}}\right) \times \left[ 1- \frac{1}{2} Q\left( \sqrt{\frac{d^2}{\sigma^2_{I}}}\right)\right]\nonumber\\
	 \label{4-qamBER} \nonumber
	 	 \end{align}
	 \normalsize
	 \vspace{-.5cm}	 

	 \noindent where $ {\frac{d^2}{\sigma^2_{I}}}$ is SINR as mentioned in \eqref{proberror}.
	 
%		 \vspace{-.6cm}

%\bibliographystyle{IEEEtran}
%\bibliography{IEEEabrv,sparse_ref}

\end{document}